%% file: asymmetric-operators.tex
\documentclass[12pt]{article} 
\usepackage{savesym,dsfont,ytableau}
\usepackage{cite}
\usepackage{picture}
\usepackage{wasysym}
\usepackage{tikz, tikz-cd}
\usepackage{amscd}
\usepackage{graphicx}
\usepackage{pifont}
\usepackage{float}
\usepackage{subfig}
\usepackage[all]{xy}
\usepackage{multicol}
\usepackage{amsfonts}
\usepackage{picture}
\usepackage{amssymb}
\savesymbol{iint}
\savesymbol{iiint}
\usepackage{amsmath}
\restoresymbol{TXF}{iint}
\restoresymbol{TXF}{iiint}
\usepackage{color}
\usepackage{clay}
\usepackage{hyperref}
\usepackage{slashed}
\hypersetup{colorlinks=true}
\hypersetup{linkcolor=black}
\hypersetup{citecolor=black}
\hypersetup{urlcolor=black}
\usepackage{setspace}
\usepackage{multirow}
\usepackage{wrapfig}
\usepackage{verbatim}
\usepackage{accents}
\numberwithin{equation}{section}
\usepackage{pdflscape}
\usepackage[T1]{fontenc}
\usepackage{varwidth}
\usepackage{amsthm}

\newcommand*{\dt}[1]{
  \accentset{\mbox{\large.}}{#1}}
\def\dot{\dt}

\newcommand{\A}{{\alpha}}
\newcommand{\B}{{\beta}}
\newcommand{\C}{{\gamma}}

\newcommand{\da}{{\dot\alpha}}
\newcommand{\db}{{\dot\beta}}
\newcommand{\dc}{{\dot\gamma}}

\newcommand{\vev}[1]{{\left< {#1} \right>}}
\newcommand{\bra}[1]{{\left< {#1} \right|}}
\newcommand{\ket}[1]{{\left| {#1} \right>}}

\newcommand*{\ditto}{--- \raisebox{-0.5ex}{\textquotedbl} ---}
\renewcommand{\arraystretch}{1.25}

\begin{document}

\date{December, 2017}

\institution{IAS}{\centerline{${}^{1}$School of Natural Sciences, Institute for Advanced Study,
Princeton, NJ, USA}}
\institution{PU}{\centerline{${}^{2}$Department of Physics, Princeton University, Princeton, NJ,
USA}}

\title{Universal Bounds on Operator Dimensions from the Average Null Energy Condition}

\authors{Clay C\'ordova\worksat{\IAS}\footnote{e-mail: {\tt claycordova@ias.edu}} and Kenan
Diab\worksat{\PU}\footnote{e-mail: {\tt kdiab@princeton.edu}}}

\abstract{We show that the average null energy condition implies novel lower bounds on the scaling
dimensions of highly-chiral primary operators in four-dimensional conformal field theories.
Denoting the spin of an operator by a pair of integers $(k,\bar{k})$ specifying the transformations
under chiral $\frak{su}(2)$ rotations, we explicitly demonstrate these new bounds for operators
transforming in $(k,0)$ and $(k,1)$ representations for sufficiently large $k$.  Based on these
calculations, along with intuition from free field theory, we conjecture that in any unitary
conformal field theory, primary local operators of spin $(k,\bar{k})$ and scaling dimension $\Delta$
satisfy $\Delta \geq \text{max}\{k,\bar{k}\}.$ If $|k-\bar{k}| > 4$, this is stronger than the
unitarity bound.}

\maketitle

\setcounter{tocdepth}{3}
\tableofcontents

\section{Introduction}
\input{intro}

\section{Constraining Operators in $(k,0)$ Representations}\label{k0-section}
\input{k0}

\section{Constraining Operators in $(k,1)$ Representations}\label{k1-section}
\input{k1}

\section*{Acknowledgements} 
We thank T.~Dumitrescu for collaboration at the initial stage of this project.  We also thank
P.~Kravchuk, J.~Maldacena, D.~Simmons-Duffin, D.~Stanford, G.~Turiaci, X.~Yin, and A.~Zhiboedov for
many helpful conversations.  C.C. is
supported by the Marvin L. Goldberger Membership at the Institute for Advanced Study,
and DOE grant de-sc0009988.

\appendix
\input{appendices}

\bibliography{asymmetric-operators}{}
\bibliographystyle{utphys}

\end{document}

%% file: intro.tex
In this paper we investigate the implications of the average null energy condition on the spectrum
of local operators in conformal field theories using the conformal collider setup of
\cite{Hofman:2008ar}.  We demonstrate that, for spinning primary operators in very chiral representations of the Lorentz group, there are universal lower bounds on scaling dimensions that are strictly stronger than those implied by the more elementary unitarity bounds of \cite{Mack:1975je, Jantzen1977}.  Based on our calculations we conjecture general formulas for these new bounds.

\subsection{Conformal Field Theories and the Unitarity Bound}
The renormalization group plays a central role in our understanding of modern quantum field theory.
At the limits of such flows one frequently finds a conformally invariant system, and therefore
conformal field theories can be viewed as the starting point for understanding general field
theories.  

The fundamental data characterizing a conformal field theory is its spectrum of local operators.
These may be organized according to their scaling dimensions $\Delta$ and spins, i.e.\ a
representation of the Lorentz group.  In four-dimensional CFTs (which are our focus here) we specify
the spin of an operator $h$ by its transformation properties under $\frak{su}(2) \oplus
\frak{su}(2)$.  This is a pair of integers $(k,\bar{k})$ specifying the number of chiral and
antichiral spinor indices carried by the operator 
\begin{equation}
h=h_{(\alpha_{1}\alpha_{2}\cdots \alpha_{k}), (\da_{1}\da_{2} \cdots \da_{\bar{k}})}~.
\end{equation}
Several familiar examples that we describe below are operators in $(k,0)$ representations, which
include (for particular $\Delta'$s) free scalars, free spinors, and field strengths of free vector
fields in the special cases $k=0,1,2$.

As is well known\cite{Mack:1975je, Jantzen1977}, in unitary CFTs all local operators transform in unitarity representations of the conformal group $\frak{so}(2,4),$
and this constrains the allowed scaling dimensions of primary local operators.  The form of these
restrictions depends on the spins $(k,\bar{k})$ as follows:
\begin{equation}
\label{unitarityboundsintro}
\Delta(k,\bar{k}) \geq \begin{cases} 0~, & k=0, \ \bar{k}=0~, \\
1+\frac{1}{2}k~, & k>0, \ \bar{k}=0~, \\
1+\frac{1}{2}\bar{k}~, & k=0, \ \bar{k}>0~, \\
2+\frac{1}{2}k+\frac{1}{2}\bar{k}~, &k>0, \ \bar{k}>0~.  
\end{cases}
\end{equation}
When the unitarity bounds above are saturated, some differential operator annihilates the local
operator.  For the case of scalars, the unitarity bound is saturated by the identity operator.  When
either $k$ or $\bar{k}$ is zero, the unitarity bound is saturated by free fields that
obey an equation of motion.  Finally, when both $k$ and $\bar{k}$ are nonzero, the unitarity bound
is saturated by conserved currents. These null states are summarized in Table \ref{nulltab}.
\begin{table}
\begin{center}
{\def\arraystretch{1.5}
\begin{tabular}{| l | l |}
\hline
Representation & Null state \\
\hline
$k = 0$, $\bar{k} = 0$ & $\partial^{\db_{1}\B_1} h$ \\
\hline
$k > 0$, $\bar{k} = 0$ & $\partial^{\db_{1}\B_1} h_{\B_1\cdots \B_{k}}$ \\
\hline
$k = 0$, $\bar{k} > 0$ & $\partial^{\db_1\B_1} h_{\db_1\cdots\db_{\bar{k}}}$ \\
\hline
$k > 0$, $\bar{k} > 0$ & $\partial^{\db_1\B_1} h_{\B_1\cdots \B_{k}\db_1\cdots \db_{\bar{k}}}$\\
\hline
\end{tabular}
}
\caption{Null states of primary operators saturating a unitarity bound.}
\label{nulltab}
\end{center}
\end{table}

The bounds \eqref{unitarityboundsintro} only take into account only the most elementary constraints
of the representation theory of the conformal group.  As we will demonstrate in this paper, these bounds
can in general be strengthened using ideas of current algebra.  A hallmark of local field theories
is that symmetry generators are obtained from integrals of local current operators.  In the case of
the conformal algebra, this operator is the energy-momentum tensor $T$.  This local operator exists
in all conformal field theories and its Ward identities encode the quantum numbers of $h$
in three-point functions of the form $\langle T hh^{\dagger}\rangle.$ Such three-point functions are our primary objects of interest. Our basic technique throughout this
work will be to constrain these three-point functions, and thereby exclude the existence of certain
local operators.

\subsection{Hints from Weinberg-Witten and Free Field Theory}\label{ww-intuition}

The fact that certain unitary representations of the conformal group listed in
\eqref{unitarityboundsintro} are incompatible with the existence of a local energy-momentum tensor
is well-known in a different guise via the Weinberg-Witten theorem \cite{Weinberg:1980kq}.  Consider a local operator in a $(k,0)$ representation of the Lorentz group (identical remarks apply for
$(0,\bar{k})$) which saturates the unitarity bound.  As described above, such an operator is a free
field satisfying an equation of motion.  In particular, when it acts on the vacuum it creates a
single massless particle with helicity $k/2$.  Such single particle states with $k>2$ are forbidden in any local
field theory with a well-defined energy-momentum tensor.  We rederive this simple result in conformal field theory language in section
\ref{ww-rederivation}.  

One of the lessons that one might draw from the Weinberg-Witten theorem is that there is a tension
between the existence of a local energy-momentum tensor, and very chiral local operators i.e.\ those
where there is a large difference between the spins $k$ and $\bar{k}$.  We can get further hints to
toward this idea by looking in more detail at the local operator spectrum of free field theories.
In four dimensions, the local operators in free field theories are constructed from polynomials in
the following basic objects:\footnote{Note that we can consider free field theories with multiple
species of a given spin, and hence below we do not require that $\xi_{\da}$ is the complex conjugate
of $\psi_{\alpha},$ nor that $G_{\da\db}$ is the complex conjugate of $F_{\alpha \beta}$.}
\begin{equation}
\label{ingredients}
\begin{tabular}{cccccccccccccccc}
$\varphi$&&&$\psi_{\alpha}$&&&$\xi_{\da}$&&&$F_{\alpha\beta}$&&&$G_{\da\db}$&&&$\partial_{\alpha\da}$\\
$\Delta=1$&&&$\Delta=3/2$&&&$\Delta=3/2$&&&$\Delta=2$&&&$\Delta=2$&&&$\Delta=1$
\end{tabular}
\end{equation}
Using these ingredients, we can attempt to build conserved currents.  As described in
\eqref{unitarityboundsintro}, these are operators that carry spinor indices of both chiralities.  It
is straightforward to see that any such operator takes the form a bilinear in the free fields
together with an arbitrary number of derivatives.  Familiar examples are symmetric conserved
currents, e.g.\
\begin{equation}
h_{(\alpha_{1}\cdots \alpha_{n})(\da_{1}\cdots\da_{n})}=\varphi \overset{\leftrightarrow}{\partial}_{\alpha_{1}\da_{1}}\overset{\leftrightarrow}{\partial}_{\alpha_{2}\da_{2}}\cdots \overset{\leftrightarrow}{\partial}_{\alpha_{n}\da_{n}} \varphi-\text{traces}~.\label{highspinintro}
\end{equation}
Here, the terminology ``symmetric" is used to indicate that these currents have no net chirality,
i.e.\ they carry an equal number of dotted and undotted indices unlike the other operators that we
describe below.  The fact that these currents occur with unbounded spin is a signature that the
theory is free: unlike the case of spins $(1,1)$ or $(2,2)$ which are compatible with non-trivial
dynamics, the Ward identities arising from currents of the form \eqref{highspinintro} with $n>2$
imply that the correlators of the energy-momentum tensor coincide with those of the free theory
\cite{Maldacena:2011jn, Alba:2015upa}. 

The classification of unitary representations in \eqref{unitarityboundsintro} reveals that there are
possible conserved currents beyond the symmetric ones described above.  In fact, such currents exist
for arbitrary spins $(k,\bar{k})$ with both $k$ and $\bar{k}$ positive.  We thus ask more generally:
which such currents may be produced using free fields? Note that if a current with spins
$(k,\bar{k})$ may be constructed, then by adding derivatives currents of spins $(k+n,\bar{k}+n)$ can be
produced for all positive $n$.  We therefore focus on the difference of the spins $|k-\bar{k}|$.
Since the Weinberg-Witten theorem constrains the spins of free fields to be those appearing in
\eqref{ingredients}, we conclude that in free field theory the net chirality of currents is bounded
as
\begin{equation}
\label{freefieldcurrents}
\text{current}~~~h_{(\alpha_{1}\cdots \alpha_{k})(\da_{1}\cdots\da_{\bar{k}})} \in \text{Free Field Spectrum} \Longleftrightarrow |k-\bar{k}|\leq 4~.
\end{equation}

The simple observation \eqref{freefieldcurrents} suggests the question: if such chiral currents do
not exist in free field theories, do they exist in any conformal field theory?  We will argue here that the answer to this question is no.   In fact, based on our calculations we
suggest that there is a non-zero gap in the spectrum of anomalous dimensions above the unitarity
bound for all operators with spins $(k,\bar{k})$ with $|k-\bar{k}|> 4$.

\subsection{The Average Null Energy Condition}

The key tool that we use to constrain the scaling dimensions of local operators is the average
null energy condition (ANEC).  This is the averaged version of the null energy condition, which
appears as a crucial assumption in many classical theorems of general relativity
\cite{Friedman:1993ty}. 

In quantum field theory, the energy-momentum tensor $T$ is a local operator that has non-trivial
quantum fluctuations, and local energy conditions do not hold \cite{Epstein:1965zza}.  However, the
averaged form has recently been established as a theorem \cite{Hofman:2016awc, Faulkner:2016mzt,
Hartman:2016lgu}.  Thus, the non-local operator $\mathcal{E}$ defined by an integral of the
null-component of $T$ along any complete null-geodesic has a non-negative expectation value in any
state $|\rho\rangle$: 
\begin{equation}
\langle\rho|\mathcal{E}|\rho\rangle
\equiv\langle\rho|\int_{-\infty}^{\infty}d\lambda  ~T_{AB}u^A u^B|\rho\rangle \geq 0~, \label{anecintro}
\end{equation}
where $u$ is the tangent null vector to the null-geodesic parameterized by $\lambda$.
Notice that only one direction is integrated over in the above.  The average null energy
operator $\mathcal{E}$ is a function of the remaining transverse coordinates. The inequality above
means that the expectation values are non-negative for all values of these transverse coordinates.
The proofs of \eqref{anecintro} in \cite{Faulkner:2016mzt, Hartman:2016lgu} link ANEC to
causality and information-theoretic entropy inequalities.  The former shows that ANEC follows from standard axioms of Euclidean conformal field theory such as crossing symmetry and reflection positivity. The latter have recently been
strengthened to semi-local versions of ANEC \cite{Bousso:2015wca, Balakrishnan:2017bjg}.

In our application we will use the average null energy condition \eqref{anecintro} to constrain the
three-point functions $\langle T hh^{\dagger}\rangle$.  As described above, the
Ward identities of the energy-momentum tensor $T$ imply that these three-point functions contain the
data of the scaling dimension and spins of $h$.  Up to a few OPE coefficients, they are also
completely fixed by conformal symmetry, and can be produced by a variety of techniques
\cite{Osborn:1993cr, Costa:2011mg, Kravchuk:2016qvl, Cuomo:2017wme}.  For the specific case of
chiral operators of interest to us we follow \cite{Elkhidir:2014woa}.

Following the pioneering work of \cite{Hofman:2008ar}, we view an operator $h$ as
creating a localized state from the vacuum, which is subject to the inequalities \eqref{anecintro}.
These ideas are closely connected to deep inelastic scattering experiments in conformal field theory
\cite{Komargodski:2016gci}. This means that complete null integrals of the three-point functions
$\langle T hh^{\dagger}\rangle$ are non-negative.  From these bounds one deduces inequalities on OPE
coefficients \cite{Hofman:2009ug, Buchel:2009sk, deBoer:2009pn, Chowdhury:2016hjy,
Chowdhury:2017vel, Cordova:2017zej}. As we describe below, in general they also imply bounds on the
scaling dimension of $h$.  

\subsection{Calculations and Conjectures}\label{conjecture-section}

With these preliminaries we can now describe the main results of this paper.  They concern the gap
above the unitarity bound for operators with general spins.  In Section \ref{k0-section} we investigate
this gap for operators $h_{(\alpha_{1}\cdots \alpha_{k})}$ in Lorentz representation
$(k,0)$.  (Identical results can be obtained for $(0,\bar{k})$ operators.)  We parameterize the
scaling dimension as   
\begin{equation}
\Delta=1+\frac{1}{2}k+\delta~,
\end{equation}
where in the above, the unitarity bounds \eqref{unitarityboundsintro} force $\delta \geq 0$.

We compute the values of $\delta$ that are compatible with the inequalities \eqref{anecintro}
applied to the three-point function $\vev{Thh^{\dagger}}$.  For $k \leq 20$ we find that
\begin{equation}
\delta \geq \frac{1}{2}k-1~.  \label{k0gap}
\end{equation}
Note that the above only becomes stronger than the unitarity bound when $k>2$.  This is consistent
with the Weinberg-Witten theorem.  For $k>20,$ the complete calculations of the conformal collider
bounds become overly technical. In these cases, however we still establish, by looking at a subset
of the inequalities, that $\delta >0$.   Based on our results, we conjecture that \eqref{k0gap}
holds for all $k$.  

\begin{center}
\begin{varwidth}{0.9\textwidth}
\underline{\textbf{Conjecture}}:
In any unitary conformal field theory, all primary local operators in $(k,0)$ representations of the
Lorentz group have scaling dimension $\Delta \geq k$.
\end{varwidth}
\end{center}

\begin{figure}[ht]
\begin{center}
\includegraphics[width=6in]{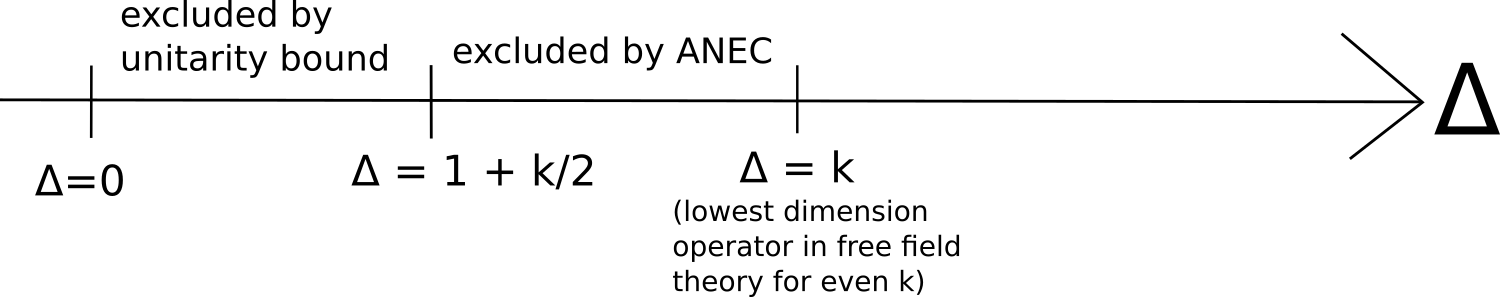}
\end{center}
\caption{Summary of the conjectured bounds for $(k,0)$ representations when $k$ is even.  The unitarity
bound sets $\Delta \ge 1 + k/2$, and the average null energy condition strengthens this to $\Delta
\ge k$.  This bound is saturated by operators constructed from free fields.}
\label{fig1}
\end{figure}

\begin{figure}[ht]
\begin{center}
\includegraphics[width=6in]{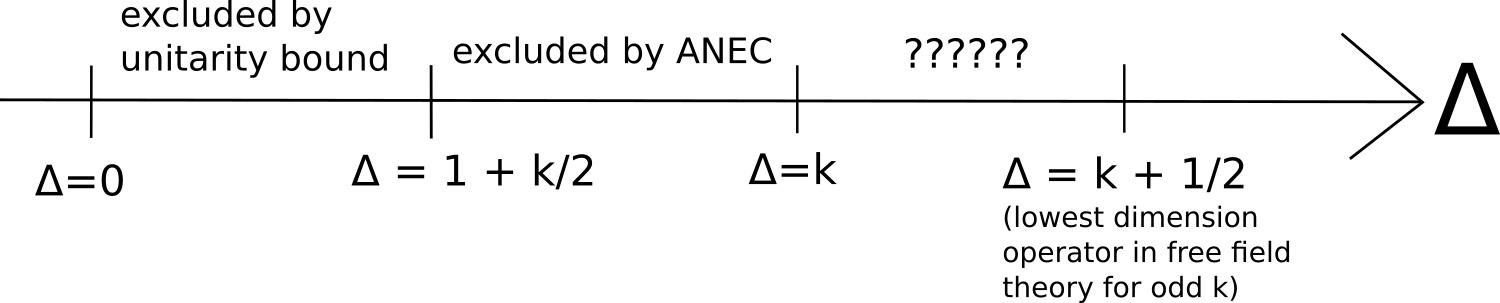}
\end{center}
\caption{Summary of the conjectured bounds for $(k,0)$ representations when $k$ is odd.  The unitarity
bound sets $\Delta \ge 1 + k/2$, and the average null energy condition strengthens this to $\Delta
\ge k$.  There is a gap of size $1/2$ between this lower bound and the lowest dimension operator of
this spin that can be constructed from free fields.}
\label{fig2}
\end{figure}

As in our discussion above, it is interesting to compare these results to the spectrum of local
operators that may explicitly be produced in free field theories.  If $k$ is even, an operator
saturating the conjectured bound may be produced by a product of free gauge field strengths, e.g.\
$F_{(\alpha_{1}\alpha_{2}}\cdots F_{\alpha_{k-1}\alpha_{k})}$, thus showing that the conjectured
bound is optimal in this case.  If instead $k$ is odd, the closest one can come to saturating the
bound in free field theory is an operator of the form $F_{(\alpha_{1}\alpha_{2}}\cdots
F_{\alpha_{k-2}\alpha_{k-1}}\psi_{\alpha_{k})}$ which has dimension $\Delta=k+1/2$.  In this case, it
is unclear whether there exist operators between this free field value and the value implied by our
conjecture.  These results are shown graphically for even $k$ in figure \ref{fig1} and for odd $k$
in figure \ref{fig2}.


In Section \ref{k1-section} we generalize our calculations to operators transforming in $(k,1)$
representations for $k\leq 7.$  We parameterize the scaling dimensions as
\begin{equation}
\Delta=\frac{5}{2}+\frac{1}{2}k+\delta~,
\end{equation} 
and as in \eqref{k0gap}, $\delta$ is non-negative by the unitarity bound.  The results of our
calculations then imply that
\begin{equation}
\delta \geq \frac{1}{2}k-\frac{5}{2}~. \label{k1gap}
\end{equation}
This is stronger than the unitarity bound precisely when $|k-1|>4$ (i.e. $k=6,7$ in our explicit
calculations). This is consistent with our discussion of chiral currents in free field theories
above.  When $k=7,$ the bound may be saturated by the free field operator $FF\partial F$
demonstrating that the result \eqref{k1gap} is optimal.  When $k=6,$ the bound cannot be saturated
by free fields and it is unclear whether operators of spin $(6,1)$ exist in the range $6\leq \Delta
< 13/2.$

Given the form of our $(k,0)$ and $(k,1)$ results, it is tempting to conjecture a general formula
for the gap in the spectrum of anomalous dimensions of general Lorentz representations implied by
the average null energy condition.  A uniform formula consistent with our calculations is stated
below. 

\begin{center}
\begin{varwidth}{0.9\textwidth}
\underline{\textbf{Conjecture}}:
In any unitary conformal field theory, all primary local operators in $(k,\bar{k})$ representations of
the Lorentz group have scaling dimension $\Delta \geq \text{max}\{k,\bar{k}\}$.
\end{varwidth}
\end{center}

As a consistency check, we note that if both $k$ and $\bar{k}$ are non-zero, our conjecture only
becomes stronger than the unitarity bound when $|k-\bar{k}|>4$ and hence is consistent with the
spectrum of free field theories.

There are a variety of possible generalizations of our work that we do not discuss.  For instance,
it would be interesting to extend these calculations to other spacetime dimensions.  In
three dimensions, all allowed unitary representations of the conformal group occur in free field
theories. However, above four dimensions there are classes of operators that do not occur in known
theories and hence it is possible that they are excluded by similar bounds. 

In another direction, the calculations performed here could be extended to superconformal field
theories.  The representation theory of these algebras is known \cite{Minwalla:1997ka, Dolan:2002zh, Dolan:2008vc, Buican:2016hpb, Cordova:2016emh}, and there are a
variety of short multiplets that do not occur in known superconformal field theories.\footnote{In
fact, even the bounds that we have derived constrain the possible short multiplets.  For instance,
in $4d$ $\mathcal{N}=1,2$ theories there are BPS chiral operators with spin $(k,0)$ and.  Our
results provide new constraints on their $U(1)_{r}$ charges.}

Finally, it would be interesting to investigate our conjectures further, and to develop a more direct
perspective on the anomalous dimensions of chiral operators.

%% file: k0.tex
In this section, we study primary operators in $(k,0)$ representations of the Lorentz group.  We denote such
an operator by $h_{\A_1\dots\A_k} = h_{(\A_1\dots\A_k)}.$  We demonstrate using the average null
energy condition that there is a gap in the spectrum of allowed scaling dimensions. This is the
simplest setting in which one can see all the relevant techniques at work.  For operators in more
complicated representations of the Lorentz group such as those investigated in section
\ref{k1-section}, there are no additional qualitative ingredients.   

As a warm-up, we begin by studying operators which behave like higher-spin free fields.  By this, we
mean operators that saturate the unitarity bound $\Delta=1+\frac{1}{2}k$.  This implies that the
operator obeys a Dirac equation
\begin{equation}
\partial^{\alpha_{1}\da}h_{\A_1\dots\A_k}=0~. \label{freefield}
\end{equation}
The Weinberg-Witten theorem \cite{Weinberg:1980kq}
implies that such higher-spin free fields are incompatible with the existence of the energy-momentum
tensor, and we rederive this result in conformal field theory language.  In this setup, it simply
states that no consistent three-point function $\vev{Thh^\dagger}$ exists. This result uses only the
constraints of conformal symmetry.  It does not require the average null energy condition.

Next, we relax the assumption that the operator $h$ is free and instead permit it to carry a general
scaling dimension $\Delta=1+\frac{1}{2}k+\delta,$ with $\delta >0$.  We explicitly solve the
conformal Ward identities and construct the most general three-point function $\vev{Thh^\dagger}.$
We then subject this correlator to the average null energy condition \eqref{anecintro}. We
find that unless the gap $\delta$ is sufficiently large, it is not possible for to satisfy these
constraints.  For $k > 2$, our lower bound is stronger than the unitarity bound.

\subsection{The  Weinberg-Witten Theorem in Conformal Field Theory}\label{ww-rederivation}

Let $h_{\A_1\dots\A_k}$ be a conformal free field.  We would like to study the three-point function
\begin{equation}
\vev{T_{(\A_1\A_2)(\da_1\da_2)}(x_1)h_{(\B_1\dots\B_k)}(x_2)h^\dagger_{(\dc_1\dots\dc_k)}(x_3)}~, \label{3ptfirst}
\end{equation}
where we have written all operators in spinor index notation with parentheses indicating
symmetrizations.  Our goal is to write the most general expression for this three-point function
compatible with Lorentz symmetry and the scaling dimensions of the operators and then impose the
constraints implied by conservation of $T$ and the Dirac equation for $h$ and $h^{\dagger}$.  

One simple way to build such three-point functions is to work in the operator product limit.  Using
the conformal group, we may move one of the operators, say $h^{\dagger},$ to infinity.  An important
advantage of this approach is that it trivializes the Ward identities from the special conformal
generators.    This means that two-point functions of descendants and primaries vanish, i.e.\
$\langle \partial^{\ell}h(0)h^{\dagger}(\infty)\rangle$ is zero unless $\ell=0$.  In this limit, we
may therefore expand the operator product $Th$ and retain only terms proportional to $h$ (and not
its descendants). 

In the OPE expansion, we find three possible structures.\footnote{This expression also arises
from employing the embedding space algorithm for constructing general three-point functions described Appendix \ref{general-3pt-section} and taking the OPE limit.} Explicitly:  
\begin{eqnarray}
T_{\alpha_{1}\alpha_{2}\da_{1}\da_{2}}(x)h_{\beta_{1}\cdots\beta_{k}}(0)&\sim &
\mathrm{Sym}_{\{\alpha_{i}\}, \{\da_{i}\},
\{\beta_{k}\}}\left[\frac{1}{x^{6}}\left(A\delta^{\chi_{1}}_{\beta_{1}}\delta^{\chi_{2}}_{\beta_{2}}x_{\alpha_{1}\da_{1}}x_{\alpha_{2}\da_{2}}+B\delta^{\chi_1}_{\A_1}\delta^{\chi_{2}}_{\beta_{2}}x_{\beta_{1}\da_{1}}x_{\alpha_{2}\da_{2}}\right.\right.
\nonumber\\ & + &
\left.\left.C\delta^{\chi_1}_{\A_1}\delta^{\chi_2}_{\A_2}x_{\B_{1}\da_{1}}x_{\B_{2}\da_{2}}\right)h_{\chi_{1}\chi_{2}\beta_{3}\cdots\beta_{k}}(0)\right]~,\label{wwth2}
\end{eqnarray} 
where in the above, $A, B,$ and $C$ are constants (OPE coefficients), and the Sym notation means
that we symmetrize over the $\A_i$, the $\da_i$, and the $\B_k$ indices independently to match the
symmetry properties of the left-hand side.

The expression \eqref{wwth2} takes into account the Lorentz symmetry and scaling dimensions of the
operators, but not the constraints of conservation of $T$ and the Dirac equation satisfied by $h$.
These are differential equations which must be satisfied by the structure appearing in the OPE limit
by adjusting the coefficients $A, B$, and $C$.  

Explicitly, to impose conservation of $T$ we demand that $\partial^{\da_1\A_1}$ annihilate the
right-hand side of \eqref{wwth2}.  To impose the Dirac equation, we shift the coordinates in the OPE
to restore the position dependence of $h,$ and then demand that $\partial^{\db\B_1}$ annihilate the
expression.   In this step the descendants of $h$ that appear on the right-hand side after taking
derivatives may be ignored since they have vanishing two-point function with $h^{\dagger}(\infty)$.

Imposing the constraints is now a straightforward calculus exercise.  In the basis of structures
given in \eqref{wwth2}, the constraints take a particularly simple form.  Each structure is independently
consistent with conservation of $T$.  Meanwhile, the implications of the equation of motion on
$h_{\alpha_{1}\cdots\alpha_{k}}$ depend on its spin $k.$
\begin{itemize} 
\item For the free scalar, $k=0$, the structures  with coefficients $B$ and $C$ do not exist, and
$\square h=0$ is automatically satisfied.  
\item For the free fermion, $k=1$, the structure with coefficient $C$ does not exist, and the Dirac
equation is satisfied when $A = 0$.  
\item For the free vector field strength, $k=2$, the Dirac equation is satisfied when $A = B = 0$.  
\item For $k > 2$, the only solution to the Dirac equation is that $A=B=C=0$.   
\end{itemize} 
In the final case enumerated above we see that $(k,0)$ free fields with $k>2$ must have vanishing
three-point function $\vev{Thh^\dagger}$.  However, as we review below, since this three-point
function encodes the Ward identities of the conformal group, it cannot vanish.  For instance,
integrals of $T$ must represent the conformal transformations on the operator $h$.  Thus,
the fact that the three-point function vanishes implies that higher-spin free fields with $k>2$ do
not exist in any theory with a conserved energy momentum tensor \cite{Weinberg:1980kq}.

\subsection{Above the Unitarity Bound: Conformal Ward Identities}\label{k0-ward}

Having established that conformal free fields of spin $(k,0)$ with $k>2$ cannot coexist with a local
stress tensor, we now relax the constraint that the field satisfy a Dirac equation, and allow the
dimension of the field to lie above the unitarity bound.  We therefore parameterize the scaling
dimension as
\begin{equation}
\Delta=1+\frac{1}{2}k+\delta~,
\end{equation}
where $\delta>0$.  Our goal is to prove that there is a gap for $\delta$, i.e.\ that it cannot
parametrically approach zero.

We first revisit the construction of the three-point function $\vev{Thh^\dagger}$, beginning with
the OPE limit \eqref{wwth2}. As mentioned above, all the structures appearing there are compatible
with conservation of the stress tensor.  Thus, for general $\delta$ where no differential operator
annihilates $h$, there are no further derivative constraints to impose.

 In general, for correlation functions involving the energy-momentum tensor $T$, there are additional
constraints from the conformal Ward identities.  These arise from the fact that the generators of
the conformal group can be expressed as integrals of the stress tensor.  More precisely, if we
contract $T$ with any conformal Killing vector $\xi$ and integrate around a small sphere containing
another operator (and no other insertions), we should obtain the action of the corresponding charge
$Q_\xi$ on that operator.  

In the context of the OPE limit relevant to our problem, this means that
\begin{equation} 
\int_{S_\epsilon^3} d\Sigma^\mu (\xi^\nu T_{\mu\nu})(x)h(0) \sim
i[Q_\xi,h](0)~,\label{e:wardint} 
\end{equation}
where  $S_\epsilon^3$ is a sphere of radius $\epsilon$ surrounding the origin.  The possible
conformal Killing vectors $\xi$ are given by: 
\begin{align} 
\textrm{Lorentz transformations:} \qquad
&(\xi_{\zeta\rho})^\nu T_{\mu\nu} = x_\rho T_{\mu\zeta} - x_\zeta T_{\mu\rho}~,
\label{e:lorentz-ward}\\ 
\textrm{Translations:} \qquad &(\xi_{\rho})^\nu T_{\mu\nu} = -T_{\mu\rho}~,\\
\textrm{Special conformal transformations:} \qquad &(\xi_{\rho})^\nu T_{\mu\nu} = 2x_{\rho}x^\nu
T_{\mu\nu} - x^2 T_{\mu\rho}~,\\ 
\textrm{Dilatations:} \qquad &\xi^\nu T_{\mu\nu} = x^\nu T_{\mu\nu}~. \label{e:lorentz-ward2}
\end{align} 
We make the dependence on the small separation $|x|$ explicit by parameterizing the
region of integration in \eqref{e:wardint} by $x^\mu = \epsilon v^\mu(x)$, where $v(x)$ is a unit
normal vector that varies over the surface of the sphere.  The measure on the surface of the $S^3$
is then 
\begin{equation} \int_{S^3_{\epsilon}} d\Sigma^\mu = \epsilon^3 \int_{S^3_{\epsilon}}d\Omega  \ v^\mu ~.
\end{equation} 

We now impose these Ward identities on the operator product expansion \eqref{wwth2}.  As described
below \eqref{3ptfirst}, on general grounds we expect no non-trivial constraints from the translation
and special-conformal Ward identities.\footnote{This can also be explicitly verified using the
conformal Killing vectors above.  The resulting integrals have an odd number of $v^{\mu}$ unit
vectors, and hence vanish when integrated over the sphere.}  However, the Lorentz transformations
and dilatations give non-trivial constraints and the desired action of the corresponding charges are
well-known.  In our spinor conventions (we mostly follow \cite{Wess:1992cp}, see Appendix
\ref{spinor-notation} for a summary) these read:
\begin{align} 
\textrm{Lorentz transformations:} \quad &
i[Q_{\xi_{\zeta\rho}},h_{\B_1\dots\B_k}](x) = \left(\sum_{i=1}^k
(\sigma_{\rho\zeta})_{\B_i}^{\chi_i} h_{\B_1\dots\B_{i-1}\chi_i\B_{i+1}\dots\B_k}(x)\right)~,\\ 
\textrm{Dilatations:} \quad & i[Q_\xi,h](x) = \Delta_hh(x)~.
\end{align} 
Performing the integrations on \eqref{wwth2} is straightforward, and we find that the above is equivalent to
\begin{equation}
B = \frac{4-2k+4\delta}{\pi^2} - 6A~, \hspace{.5in} C = \frac{-4+4k-4\delta}{\pi^2} + 6A~.  \label{e:k0-ope} 
\end{equation}
As a consistency check, one can compare these relations with the results of section
\ref{ww-rederivation} and note that these Ward identity constraints are compatible with the Dirac
equation for $k\leq 2$.  More generally, our calculation shows that for any $k$ and any
$\delta>0$, there exists a three-point function $\vev{Thh^\dagger}$ that is consistent with
conservation of $T$ and all conformal Ward identities. 

\subsection{Above the Unitarity Bound: Average Null Energy Condition}\label{k0-anec-section}

Having constructed the three-point function $\vev{T h h^\dagger}$ we now subject it to the
constraints of the average null energy condition \eqref{anecintro}.  We will compute the expectation
value of $\mathcal{E}$ in a state of definite energy $q$ and zero spatial momentum $\ket{\rho} \sim
h^\dagger(q)\ket{0}$.  The non-negativity of this expectation value will impose constraints on the
three-point function $\vev{T h h^\dagger}$.  Specifically, it will imply inequalities on the OPE
coefficients $A, B, $ and $C$, and hence via the Ward identities \eqref{e:k0-ope}, constrain the gap
$\delta$.

The analysis proceeds via the conformal collider setup of \cite{Hofman:2008ar}.  We apply a
conformal transformation to send the energy operator $\mathcal{E}$ to null infinity.  This leads to
an equivalent definition of the operator $\mathcal{E}$ that we find convenient for calculations
\cite{Zhiboedov:2013opa}.  If we define our lightcone coordinates as $x^\pm = x^0 \pm x^3$, we may
integrate over $x^-$ at $x^+\rightarrow\infty$.  In our spinor conventions (see Appendix
\ref{spinor-notation}), the null energy operator is:\footnote{In the expression below and subsequent calculations we find it convenient to use a hybrid notation involving both vector and spinor indices.  We hope that the meaning is clear from context.}
\begin{equation}
\mathcal{E} = \int_{-\infty}^\infty dx^- \lim_{x^+ \rightarrow\infty}
\frac{(x^+)^2}{16} T_{--\dot{+}\dot{+}}(x)~.
\label{e:sasha-E}
\end{equation}
Note that our choice of geodesic is completely general since rotational invariance allows us to
fix the direction in which we measure $\mathcal{E}$ to the $x^3$-direction.  Our spinor
conventions are adapted to this particular choice of geodesic.  Specifically, the $SO(2)$ rotations
around the $x^{3}$ axis are a symmetry of the problem.  We work in a basis of spinor indices where
$+,\dot{+}$ carry positive charge under this $SO(2)$ and $-\dot{-}$ carry negative charge.  Thus the
expression above for $\mathcal{E}$ is neutral.  Notice also that a parity transformation changes the sign
of the spinor indices.  We do not assume that this is a symmetry of our correlators. 

Equipped with the above definition, the process of extracting the constraints of the average null energy conditions entails
performing the following steps: 
\begin{itemize}
\item Derive the three-point function $\vev{T(x_1) h(x_2) h^\dagger(x_3)}$ at generic operator positions and
use it to generate the out-of-time order correlation function $\vev{h T h^\dagger}$.  This specific 
ordering is required to interpret the result as a one-point function of $\mathcal{E}$ in a state
created by $h^\dagger$.  
\item Using the definition of the detector \eqref{e:sasha-E}, set the indices of $T$, multiply the
expression by $(x_1^+)^2/16$ and perform the required limit and integral.
\item Fourier transform $h$ and $h^\dagger$ to give them definite energy $-q$ and $q$ respectively,
and no momentum in any spatial direction.  
\item Evaluate the result for all possible polarizations of $h$ and $h^\dagger$.  Construct the
resulting matrix of one-point functions of $\mathcal{E}$ and calculate all the eigenvalues.
\item Divide each eigenvalue by the norm of the corresponding eigenvector $\vev{h(-q)h^\dagger(q)}$.
This allows us to interpret the matrix elements as energies.
\item Demand that all these quotients are non-negative.  This yields inequalities on the OPE
coefficients $A$, $B$, and $C$.
\end{itemize}

The expression for $\vev{Thh^\dagger}$ at general operator positions can be derived in several ways.
For instance, one can apply a conformal transformation to the OPE expressions of the previous
sections to put the points at generic separations \cite{Osborn:1993cr}.  Alternatively, one can use
the techniques for constructing general three-point functions described in section
\ref{general-3pt-section} and take the OPE limit to match to the coefficients $A, B, C$ defined in
\eqref{wwth2} and thereby enforce the Ward identities.  The result is as follows.  Define $x_{ij} =
x_i-x_j$. Then, the three-point function we desire is:
\begin{multline}
\vev{T_{\A_1\A_2\da_1\da_2}(x_1)h_{\B_1\dots\B_k}(x_2)h^\dagger_{\db_1\dots\db_k}(x_3)} = 
\mathrm{Sym}_{\{\alpha_{i}\}, \{\da_{i}\}, \{\beta_{k}\}, \{\db_k\}}
\Bigg(\frac{C_h}{x_{12}^6x_{13}^6x_{23}^{2k+2\delta-4}} \times \\
\Bigg((-1)^k A \frac{x_{12}^4x_{13}^4}{x_{23}^4}  
\left( \frac{(x_{12})_{\A_1\da_1}}{x_{12}^2}- \frac{(x_{13})_{\A_1\da_1}}{x_{13}^2} \right)
\left( \frac{(x_{12})_{\A_2\da_2}}{x_{12}^2}- \frac{(x_{13})_{\A_2\da_2}}{x_{13}^2} \right)
\prod_{i=1}^k (x_{23})_{\B_i\db_i} \\
+ (-1)^{k-1} B \frac{x_{12}^2x_{13}^2}{x_{23}^2}  
\left( \frac{(x_{12})_{\A_1\da_1}}{x_{12}^2}- \frac{(x_{13})_{\A_1\da_1}}{x_{13}^2} \right)
(x_{12})_{\B_1\da_2}(x_{13})_{\A_2\db_1}
\prod_{i=2}^k (x_{23})_{\B_i\db_i} \\
+ (-1)^{k} C 
(x_{12})_{\B_1\da_1}(x_{13})_{\A_1\db_1}
(x_{12})_{\B_2\da_2}(x_{13})_{\A_2\db_2}
\prod_{i=3}^k (x_{23})_{\B_i\db_i}
\Bigg)
\Bigg)~,
\label{e:general-k0}
\end{multline}
where the Sym notation again means we symmetrize over the four sets of indices noted in the above
separately, and $C_h$ is the coefficient of the two-point function of $h$ when it is written in the form \eqref{e:general-2pt}.

Next we impose the proper operator ordering.  This is achieved with a particular $i\epsilon$
prescription.  The simplest way to see the correct prescription is to start in Euclidean signature
and analytically continue into Lorentzian signature.  A general Euclidean signature correlation
function of operators $h_{i}$ can be written as
\begin{align}
\bra{0}h_1(t^E_1,\vec{x}_1) \dots h_n(t^E_n,\vec{x}_n)\ket{0}
&= \bra{0}h_1(0,\vec{x}_1)e^{-H(t^E_1-t^E_2)} \dots e^{-H(t^E_{n-1}-t^E_{n})}
h_n(0,\vec{x}_n) \ket{0} ~.\label{orederingel}
\end{align}
This is expression is automatically time-ordered in that it is only well-defined when $t^{E}_{i}\geq t^{E}_{j}$ for $i<j$.

We now analytically continue by giving each Euclidean time an imaginary part proportional to any
desired Lorentzian time, $t^E_j \equiv \epsilon_j + it_j^L = i(t_j^L-i\epsilon_j)$.  Then, as long
as $\epsilon_i > \epsilon_j$ for all $i < j$, the operator ordering will be as written in
\eqref{orederingel} regardless of the values of the $t_j^L$.  By taking the $\epsilon_i \rightarrow
0$ after computing the correlation function, one obtains the Lorentzian correlator with a specified
operator ordering.

With these preliminaries we now proceed with the calculation.  To illustrate the entire process we
consider below the case where $h$ has spin $(3,0).$  A summary of our results for $(k,0)$
representations is discussed in Section \ref{general-k0-section}.

\subsubsection{Example Calculation: the $(3,0)$ Operator}

Using a translation, we write the general formula  \eqref{e:general-k0} for the three-point function
in the coordinates
\begin{equation}
x_{1}=y-i\epsilon~, \hspace{.5in}x_{2}=x-2i\epsilon~, \hspace{.5in}x_{3}=0~, \label{3oexplicitcoord}
\end{equation}
where in the above $\epsilon>0$ enforces the operator ordering.  When we perform the relevant
integrations, the $i\epsilon$ terms will tell us how to pick the appropriate contour.  After
integrating, we can take the limit $\epsilon \rightarrow 0.$

Following equation \eqref{e:sasha-E}, we set the indices of $T$ appropriately, multiply by
$(y^+)^2/16$ and take $y^+\rightarrow\infty$.  In this limit, the equations simplify.
In particular, by expanding norms in lightcone coordinates, e.g.~$y^2 = -y^+y^- + y_\perp^2$, we
find expressions such as:
\begin{align}
\lim_{y^+\rightarrow\infty} \frac{y_{-\dot{+}}}{y^2} &= \lim_{y^+\rightarrow\infty}\frac{y^+}{y^2} =
-\frac{1}{y^-} ~.
\end{align}
Note that if the numerator of this expression had different indices, the limit would evaluate to
zero, so for instance, this also implies expressions such as:
\begin{equation}
\lim_{y^+\rightarrow\infty} \frac{y_{\B_i\dot{+}}}{y^2} = -\frac{\delta_{\B_i}^-}{y^-} ~,
\end{equation}
where $\delta$ is the Kronecker delta symbol, i.e.~in the above, one only has a nonzero limit if the
index $\B_i$ takes value $-$. These identities also enable us to make the following simplification
on the parenthetical factor that appears multiple times in \eqref{e:general-k0} when $T$ is given the
appropriate indices:
\begin{equation}
\lim_{y^+\rightarrow\infty}\left(\frac{(x_{12})_{-\dot{+}}}{x_{12}^2}-\frac{(x_{13})_{-\dot{+}}}{x_{13}^2}\right)
= -\frac{x^-}{y^-(y^--x^-)}~.
\end{equation}
Applying all such identities, our intermediate result thus far is:
\begin{multline}
\lim_{y^+\rightarrow\infty}
\frac{(y^+)^2}{16}\vev{h_{\B_1\B_2\B_3}(x)T_{--\dot{+}\dot{+}}(y)h^\dagger_{\B_1\B_2\B_3}(0)} 
= \frac{C_h}{16}\mathrm{Sym}_{\{\beta_{k}\}, \{\db_k\}} \\ 
\Bigg(-A \frac{(x^-)^2x_{\B_1\db_1} x_{\B_2\db_2} x_{\B_3\db_3}}{(y^-)^3(y^--x^-)^3x^{6+2\delta}} 
+ B \frac{x^-\delta_{\B_1}^{-}\delta_{\db_1}^{\dot{+}} x_{\B_2\db_2} x_{\B_3\db_3}}{(y^--x^-)^3(y^-)^3x^{4+2\delta}}
- C \frac{\delta_{\B_1}^{-}\delta_{\db_1}^{\dot{+}}\delta_{\B_2}^{-}\delta_{\db_2}^{\dot{+}}x_{\B_3\db_3}}{(y^--x^-)^3(y^-)^3x^{2+2\delta}}
\Bigg)~.
\label{e:integrand-30}
\end{multline}

Our next task is to perform the integrals. Specifically, we must integrate along $y^-$ to produce
the average null energy operator, and we must Fourier transform the external states to give them
definite energy.\footnote{Note that these states, like all states of definite momentum, are delta
function normalized.  Therefore although we have two operators $h$ and $h^{\dagger},$ only one
Fourier transform is required.}

The integral along $y^-$ is straightforward to evaluate using the residue theorem once we restore the imaginary parts from \eqref{3oexplicitcoord}:
\begin{equation}
\int_{-\infty}^\infty dy^-\frac{1}{(y^--i\epsilon)^3(y^--x^-+i\epsilon)^3} =
\frac{-12i\pi}{(x^--2i\epsilon)^5}~.
\label{e:ym-int-30}
\end{equation}

The Fourier transformations in $x$ are more cumbersome since they must be carried out for each
polarization of the operators $h$ and $h^{\dagger}$ (i.e.\ for each possible choice of indices
$\B_i$ and $\db_i$).  This task can be simplified somewhat using the $SO(2)$ rotation symmetry
discussed below \eqref{e:sasha-E}.  Specifically, the final expression should be $SO(2)$ invariant
and therefore we can only get a nonzero result if there are as many $-$ and $\dot{-}$ spinor indices
as there are $+$ and $\dot{+}$ spinor indices.\footnote{Here we are using the fact that in our
spinor conventions the index names indicate the transformation properties under this $SO(2)$.  See
Appendix \ref{spinor-notation}.}  

The above argument tells us that we only need to consider pairs of polarizations corresponding to
conjugate components: $(h_{---}, h^\dagger_{\dot{+}\dot{+}\dot{+}})$, $(h_{--+},
h^\dagger_{\dot{-}\dot{+}\dot{+}})$, $(h_{-++}, h^\dagger_{\dot{-}\dot{-}\dot{+}})$, $(h_{+++},
h^\dagger_{\dot{-}\dot{-}\dot{-}})$.  The complete conformal collider bounds are therefore given by demanding that
$\mathcal{E}$ is non-negative when evaluated between any of these four pairs.  (In particular, these
expectation values are the eigenvalues of $\mathcal{E}.$)

To illustrate the mechanics of these Fourier transformations, we illustrate the case where $h$ and
$h^\dagger$ carry the polarizations $(h_{+++}, h^\dagger_{\dot{-}\dot{-}\dot{-}})$.  Combining
\eqref{e:integrand-30} and \eqref{e:ym-int-30} we have the following result:
\begin{align}
\langle h_{+++}(-q) \mathcal{E} h^\dagger_{\dot{-}\dot{-}\dot{-}}(q) \rangle &= 
\int d^4x \  e^{-iq(x^+ + x^-)/2} \frac{C_h}{16}
\Bigg(\frac{12i\pi A }{x^{6+2\delta}} 
-\frac{12i\pi B}{(x^-)^2x^{4+2\delta}}
+\frac{12i\pi C}{(x^-)^4x^{2+2\delta}}
\Bigg) ~.
\label{e:ym-int-30-2}
\end{align}
 It is simplest to evaluate the integrals in the transverse directions first.  We have
\begin{equation}
\int d^2x_\perp \frac{1}{(x^2)^k} = 
\int d^2x_\perp \frac{1}{(-x^+x^- +x_\perp^2)^k} = \frac{(-1)^{k-1}\pi}{(k-1)(x^+x^-)^{k-1}}~.
\label{e:xt-1}
\end{equation}
After applying this formula, one is left with residue integrals of the general form:
\begin{equation}
\int dx^+dx^- \frac{e^{(-iqx^+-iqx^-)/2}}{(x^+)^a(x^-)^b} = 
\frac{(2\pi)^2}{\Gamma(a)\Gamma(b)} i^{a+b} (-q/2)^{a+b-2}~.
\label{e:xpm-1}
\end{equation}
Substituting \eqref{e:xt-1} and \eqref{e:xpm-1} into \eqref{e:ym-int-30-2}, we finally obtain:
\begin{align}
\langle h_{+++}(-q) \mathcal{E} h^\dagger_{\dot{-}\dot{-}\dot{-}}(q) \rangle &= 
C_hA\frac{3 i \pi ^4 4^{-\delta} (-q)^{2\delta+2}}{\Gamma(\delta+2)\Gamma(\delta+3)}~.
\label{e:30-3pt-ex}
\end{align}

Our final task is to normalize by the two-point function so that the result can be interpreted as an
expectation value of $\mathcal{E}$.  Two-point functions of primary fields in conformal field theory
are known.  For a $(3,0)$ operator, we have (see~\eqref{e:general-2pt})
\begin{equation}
\vev{h_{\B_1\B_2\B_3}(x)h^\dagger_{\db_1\db_2\db_3}(0)} = \textrm{Sym}_{\{\B_i\}, \{\db_i\}}
C_h\frac{x_{\B_1\db_1}x_{\B_2\db_2}x_{\B_3\db_3}}{x^{2\Delta+3}}~,
\end{equation}
where $\Delta = \delta + 5/2$ is the conformal dimension of $h$.  In the polarization we are studying, the
same integration techniques can be used to calculate the two-point function of the state considered
above yielding
\begin{equation}
\langle h_{+++}(-q) h^\dagger_{\dot{-}\dot{-}\dot{-}}(q) \rangle = 
\int d^4x e^{-ip\cdot x} \vev{h_{+++}(x)h^\dagger_{\dot{-}\dot{-}\dot{-}}(0)} 
= -C_h\frac{i \pi^3 2^{1-2\delta} (-q)^{2\delta+1}}{\Gamma(\delta) \Gamma(\delta+4)}~.
\label{e:30-2pt}
\end{equation}
Taking the quotient of \eqref{e:30-3pt-ex} and $\eqref{e:30-2pt}$, we finally obtain the
conformal collider bound arising from this state:
\begin{equation}
\frac{\langle h_{+++}(-q) \mathcal{E} h^\dagger_{\dot{-}\dot{-}\dot{-}}(q) \rangle}{\langle
h_{+++}(-q) h^\dagger_{\dot{-}\dot{-}\dot{-}}(q) \rangle} =
qA\frac{3\pi(3+\delta)}{8\delta(1+\delta)}\ge 0~.
\end{equation}
Since the energy $q$ is positive and $\delta \ge 0$, this bound is satisfied only if $A \ge 0$. 

We can get stronger bounds by also including the other three nonvanishing matrix elements, which can
be evaluated with similar techniques.  Two complications can arise: first, the symmetrizations on
the indices are not trivial for two of the three-remaining polarizations, so more integrals have to
be done, and second, the numerators of the $x_\perp$ integrals can now involve factors of $x_\perp$.
Neither complication is difficult to handle.  The resulting system of four inequalities turns out to
be equivalent to
\begin{equation}
\delta \geq 1~, \hspace{.5in} 0\leq A\leq \frac{4 \delta ^2+6 \delta -4}{3 \pi ^2 \delta +15 \pi ^2}~. \label{e:30-gap}
\end{equation}
Notice in particular that $\delta >0.$  Therefore the unitarity bound on this class of operators
cannot be parametrically approached. This is our first example of bounds on operator dimensions from
the average null energy condition.

\subsubsection{Results for General $(k,0)$ Operators}\label{general-k0-section}

The calculations described above may be repeated for operators in $(k,0)$ representations.  Here we
summarize the results of our investigation.  More detailed discussion can be found in Appendix
\ref{hm-matrix-elements}.  

In general, it is difficult to analytically compute all bounds implied by the average null energy
condition simply because there are many polarizations, each of which yields an independent
inequality.  However, we can at least demand that for any particular polarization of $h$ the one
point function of $\mathcal{E}$ should be non-negative.  The polarizations that are easiest to study
are those with largest positive/negative $SO(2)$ charge, which we call the ``extremal
polarizations''.  Analogously, we can consider the next-to-extremal polarizations, which have second
largest charges.  We summarize their implications below. 
\begin{itemize}

\item The two extremal polarizations produce a system of inequalities equivalent to
\begin{equation}
\delta \geq 1~, \hspace{.5in} A\geq 0~.
\end{equation} 
Thus, for all $k>2$, the unitarity bound ($\delta=0$) cannot be saturated.

\item Including the two next-to-extremal polarizations strengthens the bound on $\delta$.  The
result depends on $k$ as follows.
\begin{equation}
k\leq 20:~~\delta \ge \frac{1}{2}k-1~, \hspace{.5in} k\geq 21:~~\delta \ge \frac{7k-6}{k-6}~. \label{bullet2}
\end{equation}
Including the next-to-next-to extermal polarizations strengthens the bound further for $k\geq21$
(but not up to $\delta \ge \frac{1}{2}k-1$) and does not alter the result for $k\leq 20$.

\item For $k$ even and $k\leq 20$ it is clear that the bound \eqref{bullet2} is optimal since it can
be saturated by the operator $F_{(\alpha_{1}\alpha_{2}}F_{\alpha_{3}\alpha_{4}}\cdots
F_{\alpha_{k-1}\alpha_{k})},$ where $F_{\alpha \beta}$ is a free gauge field strength.  For odd $k<
20,$ the closest one can get to saturating the bound in free field theory is  $\Delta=k+1/2$ via the
operator $F_{(\alpha_{1}\alpha_{2}}F_{\alpha_{3}\alpha_{4}}\cdots
F_{\alpha_{k-2}\alpha_{k-1}}\psi_{\alpha_{k})},$ where $\psi_{\alpha}$ is a free fermion.

\item For the specific case $k=3$ described above, the conformal collider bound is in fact stronger
than our conjectured bound and yields $\Delta(3,0) \geq 7/2$, matching the expectation from
free field theory.  However, we explicitly checked that for $k=5$, including all polarizations does
not strengthen the bound beyond \eqref{bullet2} (see Appendix \ref{hm-matrix-elements}).
\end{itemize}

Based on the evidence stated above, we conjecture that our results for $k\leq 20$ in fact hold for all $k$.
\begin{center}
\begin{varwidth}{0.9\textwidth}
\underline{\textbf{Conjecture}}:
In any unitary conformal field theory, all primary local operators in $(k,0)$ representations of the
Lorentz group have scaling dimension $\Delta \geq k$.
\end{varwidth}
\end{center}

%% file: k1.tex
In this section we take the first steps towards generalizing the ANEC bounds to operators in more
general representations of the Lorentz group.  Specifically, we study operators in $(k,1)$
representations.  We parameterize their scaling dimensions as
\begin{equation}
\Delta=\frac{5}{2}+\frac{1}{2}k+\delta~,
\end{equation}
where as usual, $\delta \geq 0$ by the unitarity bound.  As in the previous section, our goal is to
constrain $\delta$.  Although the analysis involves no additional conceptual ingredients, the
calculations involved are technically more challenging.  This section contains a summary of our
results.  Additional material is presented in Appendix \ref{general-3pt-section} and in the attached
{\tt Mathematica} files described in Appendix  \ref{appendixD}.

As reviewed in the introduction, when the unitarity bound is saturated $(k,1)$ representations obey a
conservation equation:
\begin{equation}
\delta=0\Longrightarrow \partial^{\db\B}h_{\B\A_2\dots\A_k\db} = 0~.
\end{equation}
Currents of this type may be explicitly produced using free fields if $k\leq 5.$  Thus, we cannot
expect anything interesting to occur in the conformal collider calculation until $k$ rises above
this range.\footnote{In fact, we  benchmarked our conformal collider calculations against
this case.  As expected we determined bounds on OPE coefficients but not the dimension $\delta$.}

Our first task is to explicitly construct the three-point function $\vev{Thh^{\dagger}}$ including
the constraints of conservation of $T$ and the conformal Ward identities.  We do this explicitly in
Appendix \ref{general-3pt-section} for $k\leq 7$ following the method of \cite{Elkhidir:2014woa} by
systematically constructing all possible conformal invariants of the correct scaling and imposing
the derivative and integral constraints.  Our results are given in Tables
\ref{tab:11}-\ref{tab:k1short} of appendix \ref{k1-explicit}.  In the end, one finds that for short
representations ($\delta=0$) the three-point function $\vev{Thh^{\dagger}}$ is specified by $\delta$
and two additional OPE coefficients, while for long representations ($\delta>0)$ there are four
additional OPE coefficients.  Hence, unlike the $(k,0)$ case, kinematical considerations do not
place constraints on operators with these spins.  Indeed, a completely consistent three-point
function $\vev{Thh^{\dagger}}$ exists for all $\delta \geq0$.  We therefore turn to an analysis of
the average null energy condition and its implications.

As compared to our calculation of $\mathcal{E}$ in section \ref{k0-section} one new qualitative
feature that appears is that the conformal collider matrix in the space of polarizations for the
operator $h$ has a more interesting structure.  Specifically, the matrix of energy one-point
functions, expressed in basis of polarizations of definite indices of $h$, is no longer diagonal
since there are multiple states that have the same $SO(2)$ charge.  Except for the extremal
polarizations $h_{-\dot{-}\dots\dot{-}}$ and $h_{+\dot{+}\dots\dot{+}}$, there are two states at
each possible value of the $SO(2)$ charge.  Hence, the matrix is block diagonal with two-by-two
blocks except for two one-by-one blocks corresponding to the extremal polarizations.  To determine
the strongest bounds we must diagonalize each block separately and compute the one-point functions
of $\mathcal{E}$ using the eigenvectors. 

In the special case of a short representation ($\delta=0$) the conservation condition in Fourier
space reads $p \cdot h = 0$.  Thus, not all polarizations of this operator are physical.
Specifically, there are $k+2$ physical polarizations obtained by starting from the $2(k+1)$ possible
values for the indices of $h$ and removing $k$ linear combinations that are not transverse to the
momentum.  In our calculation the momentum is purely in the time direction, and the preceding means
that bras and kets created by the combination $h_{x -\dot{+}} + h_{x +\dot{-}}$, where $x$ is any
multi-index, are unphysical and hence are null vectors of $\mathcal{E}$. This implies, for instance,
that one of the eigenvalues in every two by two block has to vanish.  In fact, for $k>2$ one in
general finds more zero eigenvectors than expected by this transversality argument.  Following
\cite{Zhiboedov:2013opa} it is natural to surmise that if such currents exist in the spectrum the
theory is free.  This issue will be explored in more detail in \cite{CDtoappear}.

For the case of long operators ($\delta>0$), the effects described above do not occur and all
polarizations are physical. 

The first two cases where interesting results can occur are for operators with spin $(6,1)$ and
$(7,1).$  In both these cases we computed all matrix elements of $\mathcal{E}$ analytically.  Our
findings are itemized below.

\begin{itemize}
\item  In the special case of short  $(6,1)$ and $(7,1)$ operators that saturate the unitarity bound
one finds that the $\mathcal{E}$ matrix has four non-zero eigenvalues controlled by the two free OPE
coefficients $\bar{c}_{1}$ and $\bar{c}_{2}$.  For instance, in the case of $(6,1)$ currents, the
$\mathcal{E}$ eigenvalues are, up to overall positive constants,\footnote{The energy $q$ is such a
constant we suppress in the following.} and after normalizing by the two-point function:
\begin{equation}
\{\bar{c}_1~~,~-\bar{c}_{1}-\frac{15}{14}\bar{c}_{2}+\frac{30}{7\pi^{2}}~~,~\bar{c}_1 +
\frac{15}{13}\bar{c}_2 - \frac{450}{91\pi^{2}}~~,~\bar{c}_1 + \frac{3}{2}\bar{c}_2\}~.
\end{equation}
$\mathcal{E}$ is positive semidefinite if each of the above is non-negative, and it is
straightforward to check that this system is inconsistent.  Thus it is not possible to saturate the
unitarity bound for operators in the $(6,1)$ representation.  Similarly, we find that it is not
possible to saturate the unitarity bound for $(7,1)$ operators.

\item When $\delta>0$ the matrix elements of $\mathcal{E}$ depend on four OPE coefficients and is
much more complicated.  (The full analytic form of this matrix for the case of operators in $(6,1)$
and $(7,1)$ representations is given in {\tt Mathematica} files included with the submission of this
paper.  See Appendix \ref{appendixD} for more details.) As a result of their complexity, we were
only able to analyze the constraints numerically.  Nevertheless we find sharp bounds.  Specifically,
to high numerical precision we find that 
\begin{equation}
h \in (6,1)\Longrightarrow \delta\geq 1/2~, \hspace{.5in}h \in (7,1)\Longrightarrow \delta\geq 1~. \label{numericdiscuss}
\end{equation}

\item In the case of $(7,1)$ operators, one can saturate the inequality \eqref{numericdiscuss} with
a free field operator of the form $FF\partial F$ with $F$ a free gauge field strength.  In the case of
$(6,1)$ operators the closest one can get to saturating the bound in free field theory is $\delta=1$
with an operator of the form $FF\partial \psi$ where $\psi $ is a free fermion.  \end{itemize}

These calculations motivate our more general conjecture concerning the allowed scaling
dimensions of general operators.
\begin{center}
\begin{varwidth}{0.9\textwidth}
\underline{\textbf{Conjecture}}:
In any unitary conformal field theory, all primary local operators in $(k,\bar{k})$ representations of
the Lorentz group have scaling dimension $\Delta \geq \text{max}\{k,\bar{k}\}$.
\end{varwidth}
\end{center}
We hope to investigate these ideas further in future work.

%% file: appendices.tex
\section{Spinor Notation}\label{spinor-notation}
In this paper, we follow the conventions of \cite{Wess:1992cp} with one exception: our labels for
the names of the spinor indices are nonstandard.  This is because in our calculation of the
conformal collider bounds, we give our states momentum only in the time direction, and we point the
detector in the $3$ direction.  Hence, it is useful to have a notation for the spinor components
that makes manifest the residual $SO(2)$ symmetry corresponding to rotations in the $(1,2)$ plane.
The lightcone coordinates in the $(0,3)$ plane are:
\begin{align}
y^\pm &= y^0 \pm y^3~.
\end{align}
The dictionary between vector and spinor indices is, in the conventions of Wess and Bagger:
\begin{equation}
y_\mu \sigma^\mu_{\A\da} = 
\left( {\begin{array}{cc}
   y_{1\dot{1}} & y_{1\dot{2}} \\
   y_{2\dot{1}} & y_{2\dot{2}} \\
  \end{array} } \right) =
\left( {\begin{array}{cc}
   -y_0 + y_3 & y_1 - iy_2 \\
   y_1+iy_2 & -y_0-y_3 \\
  \end{array} } \right) =
\left( {\begin{array}{cc}
   y^0 + y^3 & y^1 - iy^2 \\
   y^1+iy^2 & y^0-y^3 \\
  \end{array} } \right)~.
\end{equation}
A clockwise (positive) rotation by $\theta$ in the $(1,2)$ plane leaves $y_{1\dot{1}}$ and
$y_{2\dot{2}}$ invariant but rotates $y_{1\dot{2}}$ by $e^{-i\theta}$ and $y_{2\dot{1}}$ by
$e^{i\theta}$.  So we will give the spinor components the following new names:
\begin{equation}
y_\mu \sigma^\mu_{\A\da} = 
\left( {\begin{array}{cc}
   y_{-\dot{+}} & y_{-\dot{-}} \\
   y_{+\dot{+}} & y_{+\dot{-}} \\
  \end{array} } \right)~.
\label{e:spinor-indices}
\end{equation}
Thus, for example, we have:
\begin{align}
y_{1\dot{1}} &\equiv y_{-\dot{+}} = -y_- = y^+~, \\
y_{2\dot{2}} &\equiv y_{+\dot{-}} = -y_+ = y^-~, \\
T_{11\dot{1}\dot{1}} &\equiv T_{--\dot{+}\dot{+}} = T_{--} = T^{++}~,
\end{align}
and so on.  The third equation tells us that in our notation the null energy density $T_{--}$ in spinor
indices is $T_{--\dot{+}\dot{+}}$.

\section{Constructing and Constraining $\vev{Thh^\dagger}$}\label{appendixB}
This appendix is divided into two sections.  First, in \ref{general-3pt-section}, we will
describe how to construct the most general three-point functions of the type $\vev{Thh^\dagger}$
consistent with conformal symmetry and how one can impose various consistency conditions on these
functions in a certain OPE limit.  Then, in \ref{explicit-expressions}, we give the results
of this procedure for both short and long operators that transform in either $(k,0)$ or $(k,1)$
representations for arbitrary $k$.

\subsection{General Properties of Three-Point Functions}\label{general-3pt-section}
\subsubsection{Conformal Building Blocks for Three-Point Functions}\label{buildingblocks}
Three-point functions in conformal field theory are completely fixed by conformal symmetry up to a
set of constants because there are no cross-ratios one can write with only three points.  In four
dimensions, one can write the three-point function of generic operators $h_i$ as the
product of (a) a scalar kinematical factor $\mathcal{K}$, and (b) a linear combination of
independent tensors $\mathcal{T}_i$ that depend only on the spins of the operators.  The only
freedom is in the coefficients $c_i$ that multiply the $\mathcal{T}_i$.  
\begin{equation}
\vev{h_1(x_1)h_2(x_2)h_3(x_3)} = \mathcal{K}(x_1,x_2,x_3)\sum
c_i\mathcal{T}_i(x_1,x_2,x_3)~.
\end{equation}

The task of determining the possible $\mathcal{T}_i$ for generic three-point functions was carried
out in \cite{Elkhidir:2014woa}.  In that paper, it was proven using the embedding space formalism
that the $\mathcal{T}_i$ can be systematically generated from certain elementary ``building
blocks''.  The construction is as follows:

Let $h_i$, $i \in \{1, 2, 3\}$, be primary operators of conformal dimension $\Delta_i$ 
living in $(k_i,\bar{k}_i)$ representations, i.e.~each $h_i$ that has $k_i$ completely symmetric
undotted indices and $\bar{k}_i$ completely symmetric dotted indices.  The spins of the $h_i$ are
related to the representation labels by $s_i = (k_i+\bar{k}_i)/2$.  Let us write the index structure
of the three operators as follows:
\begin{equation}
(h_1)_{\A_1\dots\A_{k_1}\da_1\dots\da_{\bar{k}_1}}~, \quad\quad
(h_2)_{\B_1\dots\B_{k_2}\db_1\dots\db_{\bar{k}_2}}~, \quad\quad
(h_3)_{\C_1\dots\C_{k_3}\dc_1\dots\dc_{\bar{k}_3}}~. \label{e:hindices}
\end{equation}
Then, the kinematical factor is:
\begin{equation}
\mathcal{K} =
\frac{1}{
x_{12}^{(\Delta_1+s_1)+(\Delta_2+s_2)-(\Delta_3+s_3)}
x_{13}^{(\Delta_1+s_1)+(\Delta_3+s_3)-(\Delta_2+s_2)}
x_{23}^{(\Delta_2+s_2)+(\Delta_3+s_3)-(\Delta_1+s_1)}}~.
\label{e:kin}
\end{equation}
In our problem, we are not interested in the most general operators.  We will ultimately wish to
take $h_1$ to be the stress tensor $T_{\A_1\A_2\da_1\da_2}$, $h_2$ to be an operator $h$ in the
$(k,\bar{k})$ representation $h_{\B_1\dots\B_k\db_1\dots\db_{\bar{k}}}$, and $h_3$ to be its complex conjugate
$h^\dagger_{\C_1\dots\C_{\bar{k}}\dc_1\dots\dc_k}$, which transforms in the $(\bar{k},k)$ representation.
Then, $\Delta_1 = 4$, $s_1 = 2$, $\Delta_2 = \Delta_3 = 2 + \frac{k+\bar{k}}{2}$, and $s_2 = s_3 =
(k+\bar{k})/2$.
The kinematical factor then reduces to:
\begin{equation}
\mathcal{K} =
\frac{1}{x_{12}^{6} x_{13}^{6} x_{23}^{k+\bar{k}+2\Delta_h-6}}~.
\end{equation}
A similar simplification occurs for the tensor structures $\mathcal{T}_i$ of such correlators.  Only
a subset
of the most general set of building blocks are relevant.  We define this subset\footnote{
The building blocks not included in the following are the asymmetric blocks $K_{i,jk}$ and
$\bar{K}_{i,jk}$, which we have not defined.  Each $K$ carries two undotted indices, and each
$\bar{K}$ carries two dotted indices.  It turns out that there is a relation that reduces the
product of any $K$ with any $\bar{K}$ to a sum of products of $I_{ij}$ and $J_i$ tensors.  This
means that each tensor structure can be written in such a way that it contains either $K$'s or
$\bar{K}$'s, but never both.  Since the $I_{ij}$ and $J_i$ each have one dotted and one undotted
index, this implies that neither $K$ nor $\bar{K}$ can appear in a three-point function that has an
equal number of total undotted indices and dotted indices (i.e.~$k_1+k_2+k_3 =
\bar{k}_1+\bar{k}_2+\bar{k}_3$).  As mentioned, all the correlation functions we study are of this
type.  If one wanted to study correlation functions that did not have this property, however, one
would need to account for tensor structures that involve $K$ or $\bar{K}$ tensors.}
(differently from \cite{Elkhidir:2014woa}\footnote{Relative to \cite{Elkhidir:2014woa}, we have
defined the $I_{ij}$ and $J_i$ tensors to be the values one obtains after projection from six to
four dimensions instead of the six-dimensional expression.  Also, we have added a minus sign to
$I_{ij}$ for $i < j$ to simplify OPE limit expressions.  Finally, our definition of the $J_i$
differs by a factor of $\pm 1/2$, and we do not distinguish between $J_{i,jk}$ and $J_{i,kj}$ since
they are related by a minus sign.  Specifically: we chose $J_1 \equiv J_{1,23}/2$, $J_2 \equiv
J_{2,31}/2$ and $J_3 \equiv -J_{3,12}/2$.  Again, the signs are chosen to simplify OPE limit
expressions.}) as follows:
\begin{alignat}{2}
\nonumber &I_{12} = (x_{12})_{\B\da}~, \qquad && I_{21} = (x_{12})_{\A\db}~, \\
\nonumber &I_{13} = (x_{13})_{\C\da}~, \qquad && I_{31} = (x_{13})_{\A\dc}~,  \\
\nonumber &I_{23} = (x_{23})_{\C\db}~, \qquad && I_{32} = (x_{23})_{\B\dc}~,  \\
\nonumber &\displaystyle J_{1} = \frac{x_{12}^2x_{13}^2}{x_{23}^2}\left(\frac{(x_{12})_{\A\da}}{x_{12}^2} -
\frac{(x_{13})_{\A\da}}{x_{13}^2}\right)~,  
\qquad &&\displaystyle J_{2} = \frac{x_{12}^2x_{23}^2}{x_{13}^2}\left(\frac{(x_{23})_{\B\db}}{x_{12}^2} -
\frac{(x_{21})_{\B\db}}{x_{13}^2}\right)~, \\ 
&\displaystyle J_{3} = \frac{x_{13}^2x_{23}^2}{x_{12}^2}\left(\frac{(x_{13})_{\C\dc}}{x_{13}^2} -
\frac{(x_{23})_{\C\dc}}{x_{23}^2}\right)~. &&
\end{alignat}
In the above expression, one should consider the names of the indices on the right-hand side to
correspond to the indices with the same names in \eqref{e:hindices}.  (The subscripts are irrelevant
since ultimately we will symmetrize all indices of the same type.)  Then, every possible tensor
structure $\mathcal{T}_i$ can be written as a product of these building blocks such that the right
number of $\A,\da,\B,\db,\C,\dc$ indices appear, as exhibited in \eqref{e:hindices}.  Then one
symmetrizes all subsets of indices which were symmetric in the original three-point function.  That
is, one should symmetrize all the $\A_i$, all the $\da_i$, etc.  So the task of writing a general
three-point function is reduced to enumerating all possible ways of combining the structures above
appropriately.  

To perform this enumeration properly (i.e.~without including redundant structures), one has to
account for the fact that these building blocks are not automatically independent.  There is a cubic
relation that reduces $J_1J_2J_3$ to sums of products of $I_{ij}$ and $J_i$ tensors where no term
contains all three $J$'s.  This means that we should not write $\mathcal{T}_i$ that have all three
$J$'s in it.

To illustrate, we give an example.  Consider the three-point function $\vev{TVV}$ of the stress
tensor and two conserved $U(1)$ currents.  That is, we are considering $V$ to be a short $(1,1)$
representation of dimension $3$: 
\begin{equation}
\vev{T_{\A_1\A_2\da_1\da_2}(x_1)V_{\B\db}(x_2)V_{\C\dc}(x_3)} = \mathcal{K}\sum c_i\mathcal{T}_i~.
\label{e:vec-ex}
\end{equation}
The kinematical factor is $\mathcal{K} = x_{12}^{-6}x_{13}^{-6}x_{23}^{-2}$.  The possible
$\mathcal{T}_i$ are:
\begin{align}
\nonumber&\mathcal{T}_1 = I_{12}I_{13}I_{21}I_{31}~, & \qquad &\mathcal{T}_2 = I_{13}I_{31}J_1J_2~, &\qquad
&\mathcal{T}_3 = I_{12}I_{23}I_{31}J_1~, \\
&\mathcal{T}_4 = I_{13}I_{21}I_{32}J_1 ~,& \qquad &\mathcal{T}_5 = I_{23}I_{32}J_1^2~, & \qquad 
&\mathcal{T}_6 = I_{12}I_{21}J_1J_3~. 
\end{align}
As one can verify by using the definitions, each of these structures contains the correct number of
indices of each type.  For instance, we can expand the first structure as follows:
\begin{equation}
\mathcal{T}_1 = I_{12} I_{13} I_{21} I_{31} =
(x_{12})_{\B\da_1}(x_{13})_{\C\da_2}(x_{12})_{\A_1\db}(x_{13})_{\A_2\dc}~.
\end{equation}
In the above, the symmetrizations on the $\A_i$ and $\da_i$ are implicit.  These symmetrizations
must be imposed by hand. The $c_i$ appearing in \eqref{e:vec-ex} will be constrained by demanding
that $T$ is conserved, $V$ is conserved, and the conformal Ward identities are satisfied.  We 
describe how this is done below.

It is also helpful to have expressions for the two-point function $\vev{h(x_1)h^\dagger(x_2)}$.  In
this case, the only allowed building blocks are $I_{12}$ and $I_{21}$, which fixes the two-point
function completely.  If $h$ transforms in the $(k,\bar{k})$ representation, we have
\begin{equation}
\vev{h_{\A_1\dots\A_k\da_1\dots\da_{\bar{k}}}(x)h^\dagger_{\B_1\dots\B_{\bar{k}}\db_1\dots\db_k}(0)} 
= \frac{C_h}{x^{2\Delta-k-\bar{k}}}\left(\prod_{i=1}^k x_{\A_i\db_i}\right)
\left(\prod_{i=1}^{\bar{k}} x_{\B_i\da_i}\right)~,
\label{e:general-2pt}
\end{equation}
where $C_{h}$ is a constant\footnote{In these conventions, it is known that unitarity implies that
$(-i)^{k+\bar{k}}C_h > 0$ \cite{Li:2014gpa}.}.

\subsubsection{Constraints in the OPE Limit}
It is conceptually obvious how to impose the constraints arising from the conservation of $T$, a
shortening condition (if applicable), and the conformal Ward identities.
In practice, however, performing these calculations with the full three-point functions constructed
in the previous section is cumbersome.  There are a large number of components that one must check,
and the integrals relevant to the conformal Ward identities are difficult to calculate.  The
calculation simplifies dramatically if one uses the conformal symmetry to send $x_1 \rightarrow x_2$
and $x_3 \rightarrow \infty$, i.e.~take the OPE limit as $T$ approaches $h$.  Conformal symmetry
guarantees that no information is lost in this limit.  In particular, we do not have to work beyond
leading order in $x_{12}$.  At lowest nonvanishing order in $x_{12}$, the building blocks reduce to
the following expressions:
\begin{alignat}{3}
\nonumber &I_{12} \mapsto (x_{12})_{\B\da}~, \qquad\qquad && I_{21} \mapsto (x_{12})_{\A\db}~,
\qquad\qquad && I_{23} \mapsto (x_{23})_{\C\db}~, \\
\nonumber &I_{32} \mapsto (x_{23})_{\B\dc}~, \qquad\qquad && I_{13} \mapsto (x_{13})_{\C\da}~,
\qquad\qquad && I_{31} \mapsto (x_{13})_{\A\dc}~, \\
&J_1 \mapsto (x_{12})_{\A\da}~, \qquad\qquad && J_2 \mapsto (x_{12})_{\B\db}~, \qquad\qquad &&
J_3 \mapsto
\displaystyle\frac{1}{x_{12}^2}((x_{13})_{\chi\dc}(x_{13})_{\C\dot{\chi}}(x_{12})^{\dot{\chi}\chi})~.
\end{alignat}
We would like to extract the part of the OPE between $T$ and $h$ that is proportional to $h$ so that
it can be contracted with the leftover $h^\dagger$.  If we take the $x_1 \rightarrow x_2$ limit of
the full tensor structures, we will obtain expressions where the two-point function between $h$ and
$h^\dagger$ has been evaluated.  We wish to ``factor out'' this two-point function to make manifest
the exact form of the OPE.  Luckily, this is a simple task, since in the $x_1 \rightarrow x_2$
limit, $x_{13} \approx x_{23}$.  This allows us to read the two-point function directly by
extracting any piece that involves $x_3$.  For example, consider the structure $\mathcal{T}_1$ that
contributes to the $\vev{TVV}$ correlator \eqref{e:vec-ex}.  Using the dictionary above, we find
that
\begin{equation}
\mathcal{K}c_1\mathcal{T}_1 \equiv \frac{c_1}{x_{12}^6x_{13}^6x_{23}^2} I_{12} I_{13} I_{21} I_{31}
\xrightarrow{x_1\rightarrow x_2} \frac{c_1}{x_{12}^6x_{23}^8}
(x_{12})_{\B\da_1}(x_{23})_{\C\da_2}(x_{12})_{\A_1\db}(x_{23})_{\A_2\dc}~.
\end{equation}
Again, we emphasize the right-hand size of the expression above does not indicate the
symmetrizations, which must be imposed by hand.  The two-point function of $V$ is given by
\eqref{e:general-2pt}:
\begin{equation}
\vev{V_{\chi\dot{\chi}}(x_2)V_{\rho\dot{\rho}}(x_3)} =
\frac{C_V}{x_{23}^8}(x_{23})_{\rho\dot{\chi}} (x_{23})_{\chi\dot{\rho}}~.
\end{equation}
Comparing the two-point function to the OPE limit of $\mathcal{KT}_1$, we can easily identify the
two-point function in the latter expression:
\begin{equation}
\mathcal{K}c_1\mathcal{T}_1 \xrightarrow{x_1\rightarrow x_2}
\frac{c_1/C_V}{x_{12}^6}(x_{12})_{\B\da_1}(x_{12})_{\A_1\db}\vev{V_{\A_2\da_2}(x_2)V_{\C\dc}(x_3)}~. 
\end{equation}
This implies that in the OPE of $T$ with $h$, the following term appears at leading order in the
$x_1\rightarrow x_2$ limit, which we will define to be $\bar{\mathcal{T}}_1$:
\begin{equation}
T_{\A_1\A_2\da_1\da_2}(x_1)V_{\B\db}(x_2) \xrightarrow{x_1\rightarrow x_2}
\frac{c_1/C_V}{x_{12}^6}(x_{12})_{\B\da_1}(x_{12})_{\A_1\db}V_{\A_2\da_2}(x_2) + \dots 
\equiv (c_1/C_V)\bar{\mathcal{T}}_1 + \dots~.
\end{equation}
With the above example as motivation, we can simplify expressions slightly in the OPE limit
by defining rescaled coefficients 
\begin{equation}
\bar{c}_i = c_i/C_h~,
\label{e:rescaling-c}
\end{equation}
so that $T(x_1)h(x_2) \xrightarrow{x_1\rightarrow x_2} \sum \bar{c}_i \bar{\mathcal{T}}_i$,
where the $\bar{\mathcal{T}}_i$ are the OPE limits of the tensor structures.

The overall effect of the OPE limit is to ``decouple'' all the $\C$ and $\dc$ indices, making the
various constraints easier to impose.  Conservation of $T$ works straightforwardly in the OPE limit.
We simply compute:
\begin{equation}
\partial_1^{\da_1\A_1}T_{\A_1\A_2\da_1\da_2}(x_1)h_{\B_1\dots\db_1\dots}(x_2) \rightarrow 
\sum \bar{c}_i \partial_1^{\da_1\A_1}\bar{\mathcal{T}}_i~,
\end{equation}
and demand that every component vanish.  There are no derivatives on $h$ here, so this procedure
essentially amounts to just taking derivatives of the $x_{ij}$ in spinor indices, which is a
completely straightforward task.  

If $h$ is short, imposing the shortening condition on $h$ is not much harder.  For instance, if $h$
is a conserved current, we compute:
\begin{equation}
\partial_2^{\db\B_1}T_{\A_1\A_2\da_1\da_2}(x_1)h_{\B_1\dots\db_1\dots}(x_2) \rightarrow 
\sum \bar{c}_i \partial_2^{\db\B_1}\bar{\mathcal{T}}_i~.
\end{equation}
Now one might worry about derivatives on $h$ since $h$ does depend on $x_2$, but these terms are
irrelevant since they will be subleading in $x_{12}$; recall that this OPE is ultimately to be
inserted into a correlation function with $h^\dagger(x_3)$, so derivatives on $h(x_2)$ act only on
factors of $x_{23}$.

Imposing the conformal Ward identities works essentially as it did in Section \ref{k0-ward}.  As mentioned
there, we would like to contract the stress tensor $T_{\mu\nu}(x_1)$ with a conformal Killing vector
$\xi^\nu$ and integrate $x_1$ over a small sphere surrounding $x_2$.  If we write $x_{12} \sim x$
and parameterize $x^\mu = \epsilon v^\mu(x)$ as before, we would like to evaluate
\begin{equation}
\epsilon^3 \int_{S_\epsilon^3} d\Omega \  v^\mu [(\xi_{\dots})^\nu T_{\mu\nu}] h(x_2)
\end{equation}
for the various conformal Killing vectors \eqref{e:lorentz-ward}-\eqref{e:lorentz-ward2}.
Just as in the $(k,0)$ case, neither translations nor special conformal transformations impose any
constraints.  
The Lorentz transformations and dilatations do contribute, however, and the charges that correspond
to them are given by the general expressions:
\begin{align}
&\textrm{Lorentz:} \quad  i[Q_\xi,h_{\B_1\dots\B_k\db_1\dots\db_l}](x_2)
= \left(\sum_{i=1}^k (\sigma_{\mu\nu})_{\B_i}^{\chi_i} 
+ \sum_{j=1}^l (\bar\sigma_{\mu\nu})_{\db_j}^{\dot{\chi}_j}\right)
h_{\chi_1\dots\chi_k\dot{\chi}_1\dots\dot{\chi}_l}(x_2)~,\\
&\textrm{Dilatations:} \quad  i[Q_\xi,h](x_2) = \Delta_hh(x_2)~.
\end{align}
So now, all that has to be done is to evaluate the integrals corresponding to Lorentz
transformations and dilatations and demand that they evaluate to the right hand side of the above
equations.\footnote{One subtlety that can arise in this task is that Schouten identities can relate two expressions that
superficially look unequal.}

\subsection{Explicit Expressions for $\vev{Thh^\dagger}$}\label{explicit-expressions}
Now, we apply the formalism of the above to the specific cases where $h$ is either in a $(k,0)$
representation or a $(k,1)$ representation for some $k$.  We will consider both short and long
representations.  In the below, the conformal dimension of the field is always named $\Delta$.  When
we work with fields of dimensionality above the unitarity bound, we always write $\Delta = \Delta_0
+ \delta$, where $\Delta_0$ is the dimension at the unitarity bound, and $\delta > 0$.  In each
case, we will list the tensor structures $\mathcal{T}_i$ and impose the constraints imposed by the
conservation of $T$, the shortening condition (if applicable), and the conformal Ward identities.
In all cases, we give the results in terms of the rescaled OPE coefficients $\bar{c}_i$ defined in
\eqref{e:rescaling-c}.

\subsubsection{$(k,0)$ Fields} 
For primary $(k,0)$ fields, the unitarity bound sets a lower bound on the conformal dimension $\Delta$:
\begin{equation}
\Delta \ge 1 + \frac{k}{2}~.
\end{equation}

When $k=0$, $h$ is a scalar field $\phi$. There is only one tensor structure: 
\begin{align}
\mathcal{T} &= J_1^2~.
\end{align}
Conservation of $T$ and the conformal Ward identity arising from Lorentz transformations are
automatic.  The only constraint arises from the dilatation Ward identity, which sets
\begin{equation}
\bar{c} = \frac{2\Delta}{3\pi^2}~.
\end{equation}
When $\Delta=1,$ $h$ is a free scalar field and the equation of motion $\square h=0$ is automatic.

When $k=1$, $h$ is a spinor field $\psi_\A$.  There are two tensor structures:  
\begin{equation}
\mathcal{T}_1 = J_1^2 I_{32}~, \qquad
\mathcal{T}_2 = J_1 I_{12} I_{31}~.
\end{equation}
For long representations, we find:
\begin{equation}
\bar{c}_1 = \frac{2\Delta}{3\pi^2}-\frac{1}{\pi^2}~, \qquad
\bar{c}_2 = \frac{2}{\pi^2}~.
\end{equation}
Short spinor representations that saturate the unitarity bound are free fields that have $\Delta = 3/2,$ and so the
first coefficient vanishes.

When $k = 2$ there are three tensor
structures:
\begin{equation}
\mathcal{T}_1 = J_1^2 I_{32}^2~, \qquad
\mathcal{T}_2 = J_1 I_{12} I_{31} I_{32}~, \qquad
\mathcal{T}_3 = I_{12}^2 I_{31}^2~.
\end{equation}
For long representations, we find:
\begin{equation}
\bar{c}_2 = -\frac{8}{\pi^2} + \frac{4\Delta}{\pi^2}-6\bar{c}_1~, \qquad
\bar{c}_3 = \frac{12}{\pi^2} - \frac{4\Delta}{\pi^2}+6\bar{c}_1~.
\end{equation}
Short $k=2$ representations are field strengths of field vector fields and satisfy $\Delta = 2$.  The Dirac equation sets $\bar{c}_1=0$ so that
\begin{equation}
\bar{c}_1 = \bar{c}_2 = 0~, \qquad
\bar{c}_3 = \frac{4}{\pi^2}~.
\end{equation}

When $k \ge 3$, there are three tensor structures which are related to the $k=2$ structures by
powers of $I_{32}$:
\begin{equation}
\mathcal{T}_1 = J_1^2 I_{32}^{k}~, \qquad
\mathcal{T}_2 = J_1 I_{12} I_{31} I_{32}^{k-1}~, \qquad
\mathcal{T}_3 = I_{12}^2 I_{31}^2 I_{32}^{k-2}~.
\end{equation}
For long representations, we find:
\begin{equation}
\bar{c}_2 = \frac{4(\Delta-k)}{\pi^2}-6\bar{c}_1~, \qquad
\bar{c}_3 = \frac{6k-4\Delta}{\pi^2}+6\bar{c}_1~. 
\end{equation}
As proven in Section \ref{k0-section}, short representations of this type are inconsistent.

\subsubsection{$(k,1)$ Fields}\label{k1-explicit}
For primary $(k,\bar{k})$ fields with $k,\ \bar{k} \ge 1$, the unitarity bound on $\Delta$ reads:
\begin{equation}
\Delta \ge 2+\frac{k+\bar{k}}{2}~.
\end{equation}
In this section, we are concerned with fields with $\bar{k} = 1$ and $k \ge 1$.

When $k=1$, $h$ is a vector field, and there are six tensor structures.  For long representations,
we find five independent linear relations among the $\bar{c}_i$.  Short representations (vector
currents) have $\Delta = 3$, and the conservation equation does not impose any additional relations.
The tensor structures and relations are given in table \ref{tab:11}.  
\begin{table}[h]
\begin{center}
\begin{tabular}{| l | l | l | l |}
\hline
$i$ & $\mathcal{T}_i$ & $\bar{c}_i$ (long) & $\bar{c}_i$ (short) \\ \hline
\hline
$1$ & $I_{12} I_{13} I_{21} I_{31}$ & $\bar{c}_1$ & \ditto \\ \hline
$2$ & $J_1 J_2 I_{13} I_{31}$ & $\frac{3}{2}\bar{c}_1$ & \ditto \\ \hline
$3$ & $J_1 I_{12} I_{23} I_{31}$ & $\frac{2}{\pi^2}-2\bar{c}_1$ & \ditto \\ \hline
$4$ & $J_1 I_{13} I_{21} I_{32}$ & $\frac{2}{\pi^2}-2\bar{c}_1$ & \ditto \\ \hline
$5$ & $J_1^2 I_{23} I_{32}$ & $\frac{2(\Delta-3)}{3\pi^2} + \bar{c}_1$ & $\bar{c}_1$ \\\hline
$6$ & $J_1 J_3 I_{12} I_{21}$ & $-\frac{3}{2}\bar{c}_1$ & \ditto \\\hline
\end{tabular}
\caption{$\vev{Thh^\dagger}$ tensor structures and relations: $(1,1)$ field.  The last two columns
contains expressions for each $\bar{c}_i$ in terms of the free coefficient $\bar{c}_1$ for long and
short $(1,1)$ representations, respectively.  Ditto marks in the short column mean that the
expression for that $\bar{c}_i$ coincides with the corresponding expression for the long
representation.}
\label{tab:11}
\end{center}
\end{table}

When $k=2$, there are nine tensor structures.  For long representations, we find six independent
linear relations among the $\bar{c}_i$.  Short representations (supercurrents) have $\Delta = 7/2$,
and the conservation equation imposes one additional relation. The tensor structures and relations
are given in table \ref{tab:21}.

\begin{table}[h]
\begin{center}
\begin{tabular}{| l | l | l | l |}
\hline
$i$ & $\mathcal{T}_i$ & $\bar{c}_i$ (long) & $\bar{c}_i$ (short) \\ \hline
\hline
$1$ & $J_2 I_{12} I_{13} I_{31}^2 $ & $\bar{c}_1$ & \ditto \\ \hline
$2$ & $I_{12}^2 I_{23} I_{31}^2 $ & $\bar{c}_2$ & \ditto \\ \hline
$3$ & $I_{12} I_{13} I_{21} I_{31} I_{32}$ & $\bar{c}_3$ & $-2\bar{c}_1-\frac{2\bar{c}_2}{3}+\frac{8}{3 \pi^2}$ \\ \hline
$4$ & $J_1 J_2 I_{13} I_{31} I_{32}$ & $\bar{c}_1+\frac{3 \bar{c}_3}{2}$ & $-2 \bar{c}_ 1-\bar{c}_
2+\frac{4}{\pi ^2}$ \\\hline
$5$ & $J_1 I_{12} I_{23} I_{31} I_{32}$ & $-2 \bar{c}_ 1-\bar{c}_ 2-2 \bar{c}_ 3+\frac{4}{\pi ^2}$ &
$2 \bar{c}_ 1+\frac{\bar{c}_ 2}{3}-\frac{4}{3 \pi ^2}$ \\\hline
$6$ & $J_1 I_{13} I_{21} I_{32}^2$ & $-\frac{5 \bar{c}_ 1}{3}-2 \bar{c}_ 3+\frac{2}{\pi ^2}$ &
$\frac{7 \bar{c}_ 1}{3}+\frac{4 \bar{c}_ 2}{3}-\frac{10}{3 \pi ^2}$ \\ \hline
$7$ & $J_1^2 I_{23} I_{32}^2$ & $\bar{c}_ 1+\frac{\bar{c}_ 2}{6}+\bar{c}_ 3+\frac{2 \Delta }{3
\pi^2}-\frac{3}{\pi ^2}$ & $-\bar{c}_ 1-\frac{\bar{c}_ 2}{2}+\frac{2}{\pi ^2}$ \\\hline
$8$ & $J_1 J_3 I_{12} I_{21} I_{32}$ & $-\bar{c}_ 1-\frac{3 \bar{c}_ 3}{2}$ & $2 \bar{c}_ 1+\bar{c}_
2-\frac{4}{\pi ^2}$ \\\hline
$9$ & $J_3 I_{12}^2 I_{21} I_{31}$ & $-\bar{c}_1$ & \ditto \\\hline 
\end{tabular}
\caption{$\vev{Thh^\dagger}$ tensor structures and relations: $(2,1)$ field.  The third column
contains expressions for each $\bar{c}_i$ in terms of the free coefficients $\bar{c}_1$,
$\bar{c}_2$, and $\bar{c}_3$ for long representations.  For short representations, $\bar{c}_3$ is no
longer free, and so each $\bar{c}_i$ can be written in terms of $\bar{c}_1$ and $\bar{c}_2$.  Ditto
marks in the short column mean that the expression for that $\bar{c}_i$ coincides with the
corresponding expression for the long representations.}
\label{tab:21}
\end{center}
\end{table}

When $k \ge 3$, there are $10$ tensor structures.  The $k>3$ structures are generated from the $k=3$
structures by multiplying each of the $k=3$ structures by $I_{32}^{k-3}$.  We did not attempt to
perform the calculation at generic $k$, but we did work out the relations for $3 \le k \le 7$.  For
long representations, we find six independent linear relations among the $\bar{c}_i$.  From
inspection, there is a clear pattern in the relations for the long representations.  We conjecture
that the pattern continues to arbitrary $k$.  Short representations satisfy a conservation
condition, which imposes two additional relations.  For $k = 4, 5, 6, 7$, the conformal dimension of
these short representations are $\Delta = 9/2, 5, 11/2, 6$, respectively.  There is no obvious
pattern in these relations, and so we simply tabulate them explicitly for these $k$ in table
\ref{tab:k1short}.  

\begin{table}[h]
\begin{center}
\begin{tabular}{| l | l | l | }
\hline
$i$ & $\mathcal{T}_i$ & $\bar{c}_i$ (long) \\ \hline \hline
$1$ & $J_2 I_{12} I_{13} I_{31}^2 I_{32}^{k-2}$ & $\bar{c}_1$ \\ \hline
$2$ & $I_{12}^2 I_{23} I_{31}^2 I_{32}^{k-2}$ & $\bar{c}_2$ \\ \hline
$3$ & $I_{12} I_{13} I_{21} I_{31} I_{32}^{k-1}$ & $\frac{2 \bar{c}_ 4}{3}-\frac{2 \bar{c}_ 1}{3}$ \\ \hline
$4$ & $J_1 J_2 I_{13} I_{31} I_{32}^{k-1}$ & $\bar{c}_4$  \\\hline
$5$ & $J_1 I_{12} I_{23} I_{31} I_{32}^{k-1}$ & $\bar{c}_5$ \\ \hline
$6$ & $J_1 I_{13} I_{21} I_{32}^k$ & $-\frac{\bar{c}_ 1}{9}+\frac{\bar{c}_ 2}{3}-\frac{8 \bar{c}_
4}{9}+\frac{\bar{c}_ 5}{3}-\frac{2(k-3)}{3 \pi ^2}$ \\\hline
$7$ & $J_1^2 I_{23} I_{32}^k$ & $\frac{\bar{c}_ 1}{9}-\frac{\bar{c}_ 2}{6}+\frac{2 \bar{c}_
4}{9}-\frac{\bar{c}_ 5}{3}+\frac{2 \Delta }{3 \pi ^2}-\frac{k+3}{3 \pi ^2}$ \\\hline
$8$ & $J_1 J_3 I_{12} I_{21} I_{32}^{k-1}$ & $-\bar{c}_4$ \\\hline
$9$ & $J_3 I_{12}^2 I_{21} I_{31} I_{32}^{k-2}$ & $-\bar{c}_1$ \\\hline
$10$ & $J_2 J_3 I_{12}^2 I_{31}^2 I_{32}^{k-3}$ & $\frac{4 \bar{c}_ 1}{3}+2 \bar{c}_ 2+\frac{8 \bar{c}_
4}{3}+2 \bar{c}_5-\frac{4k}{\pi ^2}$ \\\hline
\end{tabular}
\caption{$\vev{Thh^\dagger}$ tensor structures and conjectured relations:
$(k,1)$ field, $k \ge 3$.  The third column contains expressions for each $\bar{c}_i$ in terms of
the free coefficients $\bar{c}_1$, $\bar{c}_2$, $\bar{c}_4$, and $\bar{c}_5$.  These relations were
verified explicitly for $3 \le k \le 7$.} \label{tab:k1-conjecture}
\end{center}
\end{table}

\begin{table}[h]
\begin{center}
\begin{tabular}{| l | l | l | l | l | l |}
\hline
 & $k=3$ & $k=4$ & $k=5$ & $k=6$ & $k=7$ \\ \hline
\hline
$\bar{c}_4$ & $\frac{6}{\pi^2} -2 \bar{c}_ 1-\frac{3 \bar{c}_2}{2}$
& $\frac{8}{\pi^2} -\frac{20 \bar{c}_ 1}{9}-2 \bar{c}_2$
& $\frac{10}{\pi^2} -\frac{5 \bar{c}_ 1}{2}-\frac{5 \bar{c}_2}{2}$
& $\frac{12}{\pi^2} -\frac{14 \bar{c}_ 1}{5}-3 \bar{c}_2$
& $\frac{14}{\pi^2} -\frac{28 \bar{c}_ 1}{9}-\frac{7 \bar{c}_2}{2}$
\\\hline
$\bar{c}_5$
& $\frac{5 \bar{c}_ 1}{3}+\frac{\bar{c}_2}{2}-\frac{2}{\pi ^2}$
& $\frac{44 \bar{c}_ 1}{27}+\frac{2\bar{c}_ 2}{3}-\frac{8}{3 \pi ^2}$
& $\frac{5 \bar{c}_ 1}{3}+\frac{5\bar{c}_ 2}{6}-\frac{10}{3 \pi ^2}$
& $\frac{26 \bar{c}_ 1}{15}+\bar{c}_2-\frac{4}{\pi ^2}$
& $\frac{49 \bar{c}_ 1}{27}+\frac{7 \bar{c}_2}{6}-\frac{14}{3 \pi ^2}$
\\\hline
\end{tabular}
\caption{Expressions for $\bar{c}_4$ and $\bar{c}_5$ in terms of the free coefficients $\bar{c}_1$
and $\bar{c}_2$ implied by the shortness condition for $(k,1)$ fields, $3 \le k \le 7$}
\label{tab:k1short}
\end{center}
\end{table}

\section{Conformal Collider Inequalities for $(k,0)$ Operators}\label{hm-matrix-elements}

In this section, we tabulate the conformal collider inequalities in complete detail (including all
polarizations) for $(k,0)$ operators with $3\leq k \leq 6$. 

When $k = 3$, the full inequalities are equivalent to:
\begin{align}
\delta \geq 1~, \hspace{.5in} 
0\leq A\leq \frac{4 \delta ^2+6 \delta -4}{3 \pi ^2 \delta +15 \pi ^2}~.
\end{align}

When $k = 4$, the full inequalities are equivalent to:
\begin{align}
\left(1 \le \delta \le \xi_4 \textrm{ and } 0 \le A \le \frac{2\delta^2 + 4\delta - 6}{\pi^2\delta
+ 11\pi^2}\right)
\textrm { or } \left(\delta \ge \xi_4 \textrm { and } 0 \le A \le
\frac{4\delta^2+2\delta^2-12\delta+54}{3\pi^2\delta^2 + 6\pi^2\delta - 45\pi^2}\right)~,
\end{align}
where $\xi_4 \approx 14.596$ is the unique real solution to $x^3 - 14x^2 -5x-54 = 0$.

When $k = 5$, the full inequalities are equivalent to:
\begin{align}
\left(\frac{3}{2} \le \delta \le \xi_5 \textrm{ and } 0 \le A \le \frac{4\delta^2 + 10\delta - 24}{\pi^2\delta
+ 29\pi^2}\right)
\textrm { or } \left(\delta \ge \xi_5 \textrm { and } 0 \le A \le
\frac{3\delta^3 + 3\delta^2 - 18\delta + 72}{2\pi^2\delta^2 + 10\pi^2\delta - 42\pi^2}\right)~,
\end{align}
where $\xi_5 \approx 10.223$ is the unique real solution to $x^3 - 10x^2 +3 x -54 = 0$.

When $k = 6$, the full inequalities are equivalent to:
\begin{align}
\nonumber \left(2 \le \delta \le \xi_6 \textrm{ and } 0 \le A \le \frac{\delta^2 + 3\delta -
10}{9\pi^2}\right)\\
\nonumber \textrm { or } \left(\xi_6 \le \delta \le \xi_6' \textrm { and } 0 \le A \le
\frac{16\delta^3 + 24\delta^2 - 160\delta + 600}{9\pi^2\delta^2 + 81\pi^2\delta - 270\pi^2}\right)\\
\textrm { or } \left(\delta \ge \xi_6' \textrm { and } 0 \le A \le
\frac{3\delta^3 + 7\delta^2 + 200}{2\pi^2\delta^2 + 8\pi^2\delta - 60\pi^2}\right)~,
\end{align}
where $\xi_6 \approx 8.861$ is the unique real solution to $x^3 - 9x^2 +8 x -60 = 0$ and
$\xi_6' \approx 31.635$ is the unique real solution to $x^3 - 29x^2 -72 x -360 = 0$.

\section{Conformal Collider Inequalities for $(k,1)$ Operators}\label{appendixD}
As mentioned in the text, the conformal collider inequalities are too cumbersome to quote in the text
or analyze analytically, so we include them as \texttt{Mathematica} files.  In this section, we
briefly describe how these files are presented and how to work with them.

The file \texttt{E61} contains the full $14$ by $14$ array of $\mathcal{E}$ matrix elements in
the case where $h$ transforms in the $(6,1)$ representation, and the file \texttt{twopoint61}
contains the corresponding array of two-point functions $\vev{hh^\dagger}$.  The file \texttt{E71} and
\texttt{twopoint71} contain the analogous $16$ by $16$ arrays in the case where $h$ transforms in
the $(7,1)$ representation.

As explained in Section \ref{k1-section}, there is a basis of polarizations for $h$ and $h^\dagger$
where all of these matrices are block diagonal.  In the included files, the rows denote bra states
created by $h_{\dots}$ and the corresponding columns denote the conjugate ket states created by
$h^\dagger_{\dots}$.  The state with minimal $SO(2)$ charge $h_{-\dots-\dot{-}}$ corresponds to a
one-by-one block, as does the state with maximal $SO(2)$ charge $h^\dagger_{+ \dots +\dot{+}}$.  All
other states live in two-by-two blocks.  We write the state with minimal $SO(2)$ charge in the upper
left and the state with maximal $SO(2)$ charge in the lower right, and the two-by-two blocks
corresponding to states having charges of intermediate values are in order, decreasing in units of
two charge from upper left to lower right.  In each two-by-two block, we order the rows
$h_{-\dots\-\dot{+}}$, $h_{+\dots\dot{-}}$ so that the conjugate states labeling the columns are
$h^\dagger_{-\dots\dot{+}}, h^\dagger_{+\dots\dot{+}}$.  For example, the first three rows in the
$(6,1)$ case correspond, in order, to the states $h_{------\dot{-}}, h_{------\dot{+}},
h_{-----+\dot{-}}$ so that the first three columns correspond to the conjugate states
$h^\dagger_{+\dot{+}\dot{+}\dot{+}\dot{+}\dot{+}\dot{+}},
h^\dagger_{-\dot{+}\dot{+}\dot{+}\dot{+}\dot{+}\dot{+}},
h^\dagger_{+\dot{-}\dot{+}\dot{+}\dot{+}\dot{+}\dot{+}}$.

The files themselves are Wolfram Language expressions that can be imported using the \texttt{Get}
function, e.g.~\texttt{mat = Get["E61"]} loads the $(6,1)$ array of $\mathcal{E}$ matrix elements
into the variable \texttt{mat}, provided that the file \texttt{E61} is located in the present
working directory.